# High-throughput search for magnetic topological materials using spin-orbit spillage, machine-learning and experiments

Kamal Choudhary[1,2], Kevin F. Garrity[1], Nirmal J. Ghimire[3,4], Naween Anand[5], Francesca Tavazza[1]

1 Materials Science and Engineering Division, National Institute of Standards and Technology, Gaithersburg, MD, 20899, USA.

2 Theiss Research, La Jolla, CA 92037, USA.

3. Department of Physics and Astronomy, George Mason University, Fairfax, VA 22030, USA.

4. Quantum Science and Engineering Center, George Mason University, Fairfax, VA 22030, USA.

5. Materials Science Division, Argonne National Laboratory, Argonne, Illinois 60439, USA.

## Abstract

Magnetic topological insulators and semi-metals have a variety of properties that make them attractive for applications including spintronics and quantum computation, but very few high-quality candidate materials are known. In this work, we use systematic high-throughput density functional theory calculations to identify magnetic topological materials from the ≈40000 three-dimensional materials in the JARVIS-DFT database (https://jarvis.nist.gov/jarvisdft). First, we screen materials with net magnetic moment > 0.5 μB and spin-orbit spillage > 0.25, resulting in 25 insulating and 564 metallic candidates. The spillage acts as a signature of spin-orbit induced band-inversion. Then, we carry out calculations of Wannier charge centers, Chern numbers, anomalous Hall conductivities, surface bandstructures, and Fermi-surfaces to determine interesting topological characteristics of the screened compounds. We also train machine learning models for predicting the spillage, bandgaps, and magnetic moments of new compounds, to further accelerate the screening process. We experimentally synthesize and characterize a few candidate materials to support our theoretical predictions.

**Corresponding author**: kamal.choudhary@nist.gov

## Introduction

The interplay of topology[1,2] and electronic band structures in non-magnetic materials has led to several new material categories, most notably topological insulators (TI)[1,3], Dirac semi-metals, and broken-inversion Weyl semimetals (SM)[4,5], topological crystalline insulators[6] , nodal line semi-metals[7,8]. However, many potentially useful quantum effects[9-13], like anomalous Hall conductivity (AHC), are only possible in topological materials with broken time reversal symmetry (TRS), including exotic phases such as Chern insulators[9,14], magnetic axion insulators[9,15], and magnetic semimetals[16]. Experiments such as anomalous Hall conductivity[17], spin-Seebeck[18], spin-torque ferromagnetic resonance[19], and angle-resolved photoemission spectroscopy (ARPES)[20], Fourier transform scanning tunneling spectroscopy (FT-STS)[21], and Shubnikov de Haas (SdH) oscillations can be useful for analyzing the topological behavior. Only a few such materials are reported experimentally, and many of those materials are limited to very low temperatures or have trivial bands that overlap with the topological band features, limiting their utility. There is a significant opportunity to find more robust magnetic topological materials and to further our understanding of the underlying mechanisms leading to their topological properties.

A common feature of many topological materials classes is the presence of spin-orbit induced band inversion, where the inclusion of spin-orbit coupling in a calculation causes the character of the occupied wavefunctions at a k-point to change. Spin-orbit spillage (SOS)[22-24] is a method to measure this band inversion by comparing the wavefunctions with and without spin-orbit coupling (SOC). SOS is based on density functional theory (DFT) calculations based wavefunction analysis and has been proven to be a useful technique for finding topological materials. Previous studies[22-24] have looked at three-dimensional (3D) non-magnetic materials as well as two-dimensional (2D) materials with and without magnetism. Due to its ease of calculation, without any need for symmetry analysis or dense k-point interpolation, the SOS is an excellent tool for identifying candidate materials to many topological phases. Advantages of the spillage technique include that it can apply to materials with low or no symmetries, including disordered or defective materials, and that it can identify the fundamental driver of topological behavior, the band inversion, even if the exact topological classification a material will depend on detailed features like the exact magnetic ordering, spin-direction, or sample thickness. After identifying high spillage materials, further analysis is necessary to identify the specific topological phases that may arise from the band inversion.

Stoichiometric magnetic topological insulators (MTIs) are very rare. $MnBi_2Te_3$[25,26], an antiferromagnetic TI, is one of the most studied and well-characterized examples of a 3D MTI, and thin films of $MnBi_2Te_3$ exhibit quantized AHC[26]. Several magnetic semimetals (MSM), such as CuMnAs, $Fe_3GeTe_2$, LaCl, $EuCd_2As_2$ have been reported as well[16]. Recently there have been several efforts to systematically identify topological materials, especially for the non-magnetic systems[23,24,27-29] . The spin orbit spillage technique has been successfully used to identify thousands of 3D non-magnetic insulators, semi-metals[23] as well as 2D non-magnetic and magnetic insulators and semimetals such as $VAg(PSe_3)_2$, $ZrFeCl_6$, MnSe and $TiCl_3$[24]. Identification of MTIs and MSMs has been developed by topological quantum chemistry groups[30,31] in which wavefunction symmetry indicators are used to identify topological materials.

In this work, we screen for 3D magnetic topological insulators (MTI) and semimetals using the SOS technique. We then analyze the resulting high-spillage materials using conventional Wannier tight-binding Hamiltonian-based techniques to calculate Chern numbers, anomalous Hall conductivities, Berry-curvatures, and Fermi-surfaces, as well as to local band crossings. Starting with crystal structures optimized using the OptB88vdW[32] van der Walls functional, we first identify materials using the Perdew-Burke-Ernzerhof (PBE)[33] generalized gradient approximation (GGA) functional, and then carry out Strongly Constrained and Appropriately Normed (SCAN)[34] meta-GGA functional calculations of a subset of materials.

While our DFT-based computational screening is relatively efficient, it is still computationally expensive when applied to a set of thousands of materials. To further accelerate the identification and characterization process, we develop classification machine learning models for metals/non-metals, magnetic/non-magnetic and high-spillage/low-spillage materials, which acting together can screen topological materials in different classes. Specifically, we use JARVIS-ML based classical force-field inspired descriptors (CFID)[35] and gradient boosting decision tree (GBDT) for developing the ML models. CFID based models have been successfully been used for developing more than 25 high-accurate ML property prediction models[36]. Using this approach, we can first predict topological materials using ML, then confirm with SOS and Wannier tight-binding approaches. The selected materials can be promising for experimental synthesis and characterizations. All the data and models generated through this work are publicly distributed through JARVIS-DFT[23,24,36], JARVIS-WTB[37] and JARVIS-ML webapps[36]. We also share the computational tools and workflows developed for this work through JARVIS-Tools open access software to enhance the reproducibility and transparency of our work. As spillage is a computational screening technique for topological materials, there are many experimental techniques to delineate topological characteristics such as ARPES, SHE, and QHE. In this paper, we use some of these techniques to support the findings of spillage based two screened materials.

This paper is organized as follows: first we show the screening strategy for high-spillage magnetic materials and present statistical analysis of some of their properties. Next, we show bandstructures and k-point dependent spillage for a few example candidate materials to illustrate the strategy. After that we further analyze selected insulating and metallic band structures with Wannier tight-binding approaches. Then, we analyze the periodic table distribution trends and develop machine learning classification models to accelerate the identification processes. Finally, we show experimental characterizations of a few candidate materials.

## Results and discussion

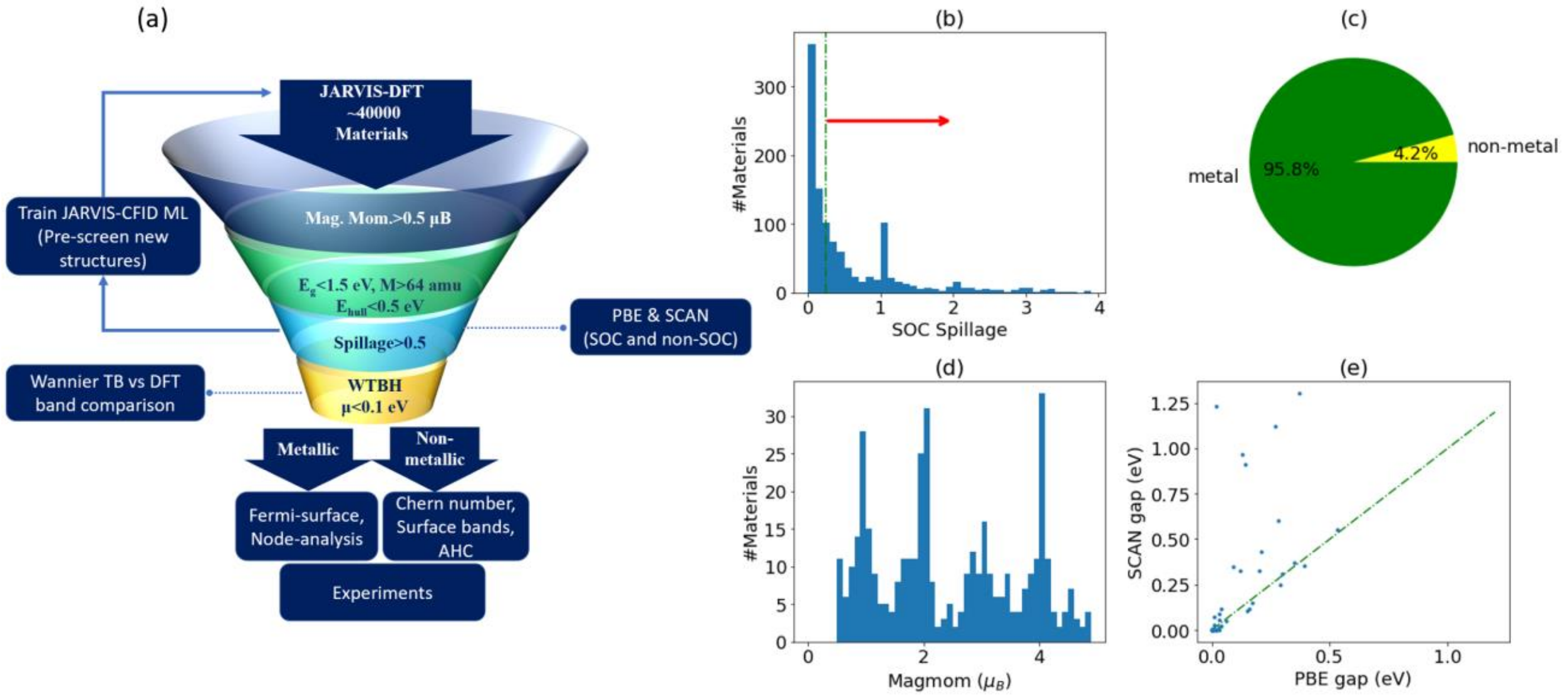

*Fig. 1 Flow-chart for screening high-spillage materials and analysis. a) flowchart for screening, b) spillage distribution analysis for all the materials under investigation, c) pie chart showing high spillage insulators and metals, d) magnetic moment distribution for high-spillage materials, e) PBE vs SCAN bandgaps.*

A flow chart for screening magnetic topological materials is shown in Fig. 1a. First, we screen for materials with net magnetic moment (>0.5 μB) in the ferromagnetic phase, which leads to 8651 candidates out of 39315 materials in the JARVIS-DFT database. Then we look for materials that are reasonably stable and are likely to display topological band inversion by screening for materials that: a) are less than 0.5 eV/atom above the convex hull[38], b) have small non-SOC bandgaps (<1.5 eV), and c) have at least one atom with high atomic mass (M>64). This results in 4734 remaining materials. We have computed the spin-orbit spillage (SOS) with PBE+SOC for 1745 materials (prioritizing the calculations of the number of atoms in unit cell less than 20). Next, we perform Wannier tight-binding Hamiltonian (WTBH) calculations with high quality (MaxDiff<0.1 eV)[37] to predict topological invariants, surface bandstructures, Fermi-surfaces, and anomalous Hall conductivity. So far, we have obtained high-quality WTBHs for 146 candidate materials. To study the effects of exchange-correlation, we run (SCAN)[34] meta-GGA functional calculations for high-spillage materials. Note that it may be difficult to carry out high dense k-point DFT calculations with SOC for thousands of materials, so after the WTBH generation, we carry our high-density k-point calculation Wannier TB models to find if the bandgap truly exists. Most of the materials studied in this work come from experimentally determined structures from the inorganic crystal structure database (ICSD) [39].

In Fig. 1b we show the spillage distribution of the materials investigated in this work. As the spillage can be related to the number of band-inverted electrons at a k-point, we observe spikes at integer numbers[22-24]. Spin-orbit coupling can also change the mixing between different orbitals, rather than pure band inversion, which results in fractional spillage amounts. As shown in Fig. 1b using a spillage threshold of 0.25 for screening eliminates 51 % of materials, leaving 25 insulating and 564 metallic candidate materials with high spillage and non-zero magnetic moment. Similarly, in our previous works for 3D non-magnetic and 2D materials[22-24], spillage technique was shown to discard more than 50 % candidates in the initial screening steps also. A material with non-zero spillage is a candidate topological material and we choose a threshold of 0.25 to narrow down the options. In Fig.1c we show the pie chart for high spillage insulating and metallic materials distribution. This suggests that magnetic topological insulators (MTI) are far rarer than semimetals. In the later sections, we discuss with examples some of the insulating and metallic high-spillage materials and characterize them using Wannier tight-binding Hamiltonian approach also. Next in Fig 1d, we observe that the magnetic moment of the systems could be up to 6 μB with mostly integer or close to integer values for the magnetic moments. Due to the large computational expense of searching for magnetic ground states, we only considered ferromagnetic spin configuration i.e., all spins of the system in a fixed direction. We expect that many of high-spillage materials that we find to be ferromagnetic may turn out to have lower energies in the anti-ferromagnetic or ferri-magnetic configurations. In Fig. 1e, we compare the bandgaps of the materials with PBE+SOC and SCAN+SOC for 65 high-spillage materials. Recently the SCAN functional has been proposed as the functional to solve the bandgap and high correlated system issues which can be important for magnetic topological materials. SCAN has been shown to predict bandgaps and magnetic moments better than LDA, LDA+U, and PBE in many cases[40-42]. We observe that SCAN+SOC bands are very close or in some cases slightly higher than PBE+SOC bandgaps for most of the materials. However, for some systems, it can be up to 10 times larger such as for $LiVH_2OF_5$ (JVASP-47705). Some of the materials that are metallic in PBE turns into insulating in SCAN predictions (for example, $LiMnAsO_4$(JVASP-55805), $Li_4Fe_3CoO_8$ (JVASP-42538)), which indicates that magnetic metals found to be high spillage using PBE may in fact be small gap topological insulators. We provide more detailed PBE vs. SCAN comparisons in the supplementary information (Table S1).

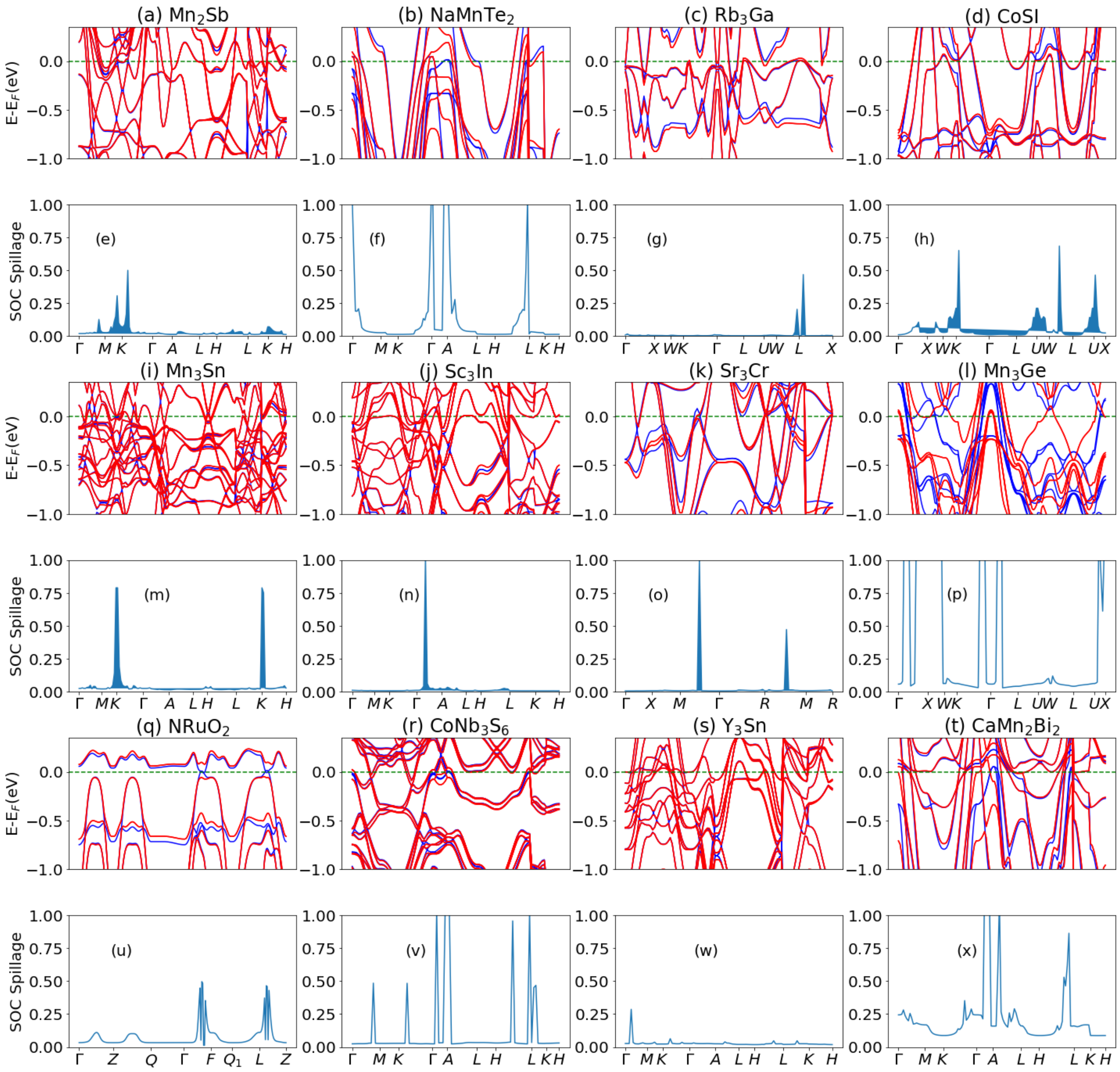

*Fig. 2 Examples of bandstructure and k-dependent spin-orbit spillage plots for a few selected candidate materials with PBE+SOC. Bandstructures are shown in a) $Mn_2Sb$ (JVASP-15693), b) $NaMnTe_2$ (JVASP-16806), c) $Rb_3Ga$ (JVASP-38248, d) CoSI (JVASP-78508), i) $Mn_3Sn$ (JVASP-18209), j) $Sc_3In$ (JVASP-17472), k) $Sr_3Cr$ (JVASP-37600), l) $Mn_3Ge$ (JVASP-78840), q) $NaRuO_2$(JVASP-8122), r) $CoNb_3S_6$ (JVASP-21459), s) $Y_3Sn$ (JVASP-37701), t) $CaMnBi_2$ (JVASP-18532). The red and blue lines show SOC and non-SOC bandstructures respectively. The k-dependent spillage is shown in (e), (f), (g), (h), (m), (n), (o), (p), (u), (v), (w) and (x) respectively.*

In Fig. 2 we show the non-spin orbit and spin-orbit bandstructures for a few screened insulating and semi-metallic systems along with corresponding spin-orbit spillage plots such as a) $Mn_2Sb$ (JVASP-15693), b) $NaMnTe_2$ (JVASP-16806), c) $Rb_3Ga$ (JVASP-38248, d) CoSI (JVASP-78508), i) $Mn_3Sn$ (JVASP-18209), j) $Sc_3In$ (JVASP-17472), k) $Sr_3Cr$(JVASP-37600), l) $Mn_3Ge$ (JVASP-78840), q) $NaRuO_2$(JVASP-8122), r) $CoNb_3S_6$ (JVASP-21459), s) $Y_3Sn$ (JVASP-37701), t) $CaMnBi_2$ (JVASP-18532). The red and blue lines show SOC and non-SOC bandstructures respectively. The k-dependent spillage is shown in (e), (f), (g), (h), (m), (n), (o), (p), (u), (v), (w) and (x) respectively. Such bandstructures and spillage plots for 11483 materials (including 2D and 3D magnetic and non-magnetic systems) are distributed through the JARVIS-DFT website along with several other materials properties such as crystal structure, heat of formation, elastic, piezoelectric, dielectric, and thermoelectric constants. In all the cases, the spillage is higher than 0.25 and the magnetic moments in the ferromagnetic configuration for these systems are more than 1 $\mu$B. The $NaRuO_2$ shows a PBE+SOC gap of 56 meV while other materials are metallic. We note that in some cases, the magnetic ordering or magnetic moment can change significantly when adding SOC to a calculation, resulting in a high spillage value without any direct relation to band inversion. Hence, it is important to further analyze the candidate materials by directly computing topological behavior, and we show examples of this analysis for $NaRuO_2$ and $Y_3Sn$ below.

In our earlier work[37], we created a database of automatically generated WTBH, which we use here to analyze topological behavior and support our findings from the spillage-based screening. The accuracy of the WTBH is evaluated based on the MaxDiff criteria[37] which compares the maximum band-energy difference between DFT and WTB on k-points within and beyond our DFT calculations k-points. We set a MaxDiff (maximum energy difference at all k-points between Wannier and DFT bands) value of 0.1 eV as the tolerance for a good-quality WTBH. Out of all the spillage-based candidate materials we observe at least 146 high of them have low MaxDiff. For the systems with high spillage and high-quality WTBH, we predict Wannier charge centers, surface bandstructures, and anomalous Hall conductivity for the insulating cases and AHC, Fermi-surfaces and node plots for the metallic cases. Our Wannier database is available at https://jarvis.nist.gov/jarviswtb/ with interactive features. We provide heat of formation, spacegroup, convex hull and other important details for each material in the corresponding webpage (such as https://www.ctcms.nist.gov/~knc6/static/JARVIS-DFT/JVASP-8122.xml)as well as in the supplementary information (Table S2) These webpages can also be downloaded as XML documents containing raw data for replotting or analysis by the users.

We identify $NaRuO_2$ as a candidate 3D Chern insulator through the above systematic screening process based on PBE+SOC and SCAN+SOC. $NaRuO_2$ is a trigonal system, belonging to $R\bar{3}m$ spacegroup. The heat of formation of the system is negative (-1.293 eV/atom) suggesting the system should be thermodynamically favorable. Also, the system has a formation energy that is 0.089 eV/atom above the convex hull, suggesting that the system is slightly unstable but in a range where is may be synthesizable, and it has in fact been synthesized experimentally[43]. We observe that this material is metallic without SOC (Fig. 2a), but as we turn on SOC, a gap opens at the B and X points, which results in high spillage of 0.56. At least 18 materials show bandgap opening due to inclusion of spin-orbit coupling. Next, we calculate the Chern number using the Wannier charge centers as shown in Fig. 3a and b. We observe gapless charge centers, indicating that the material is a 3D Chern insulator. The Chern number of four planes i.e., $k_1$=0.0; $k_1$=0.5; $k_2$=0.0; $k_2$=0.5 ($k_3$=0.0; $k_3$=0.5 and $k_2$=0.0; $k_2$=0.5 remaining the same); where $k_1$, $k_2$, $k_3$ is in fractional units is determined as -2. In Fig. 3c we see a conducting channel in the (001) surface suggesting that the material is conducting at its surface, but the bulk is insulating even though the time reversal is broken in the system. The Chern number is directly proportional to the anomalous Hall conductivity which is an experimentally measured quantity. For a 3D Chern material, AHC is calculated as $\frac{C_3 b_3 e^2}{2\pi h}$ which turns out to be 1540 ohm$^{-1}$cm$^{-1}$, which is what we find using Wannier calculation-based quantity in the Fig. 3d. In this case the AHC in Fig. 3d is quantized which can be leveraged for precise quantum control from the perspective of building devices. In addition, we analyzed this material using SCAN+SOC, and we find that the band structure is very similar to the PBE+SOC result, and the topological properties are the same (see the supplementary information Fig. S1).

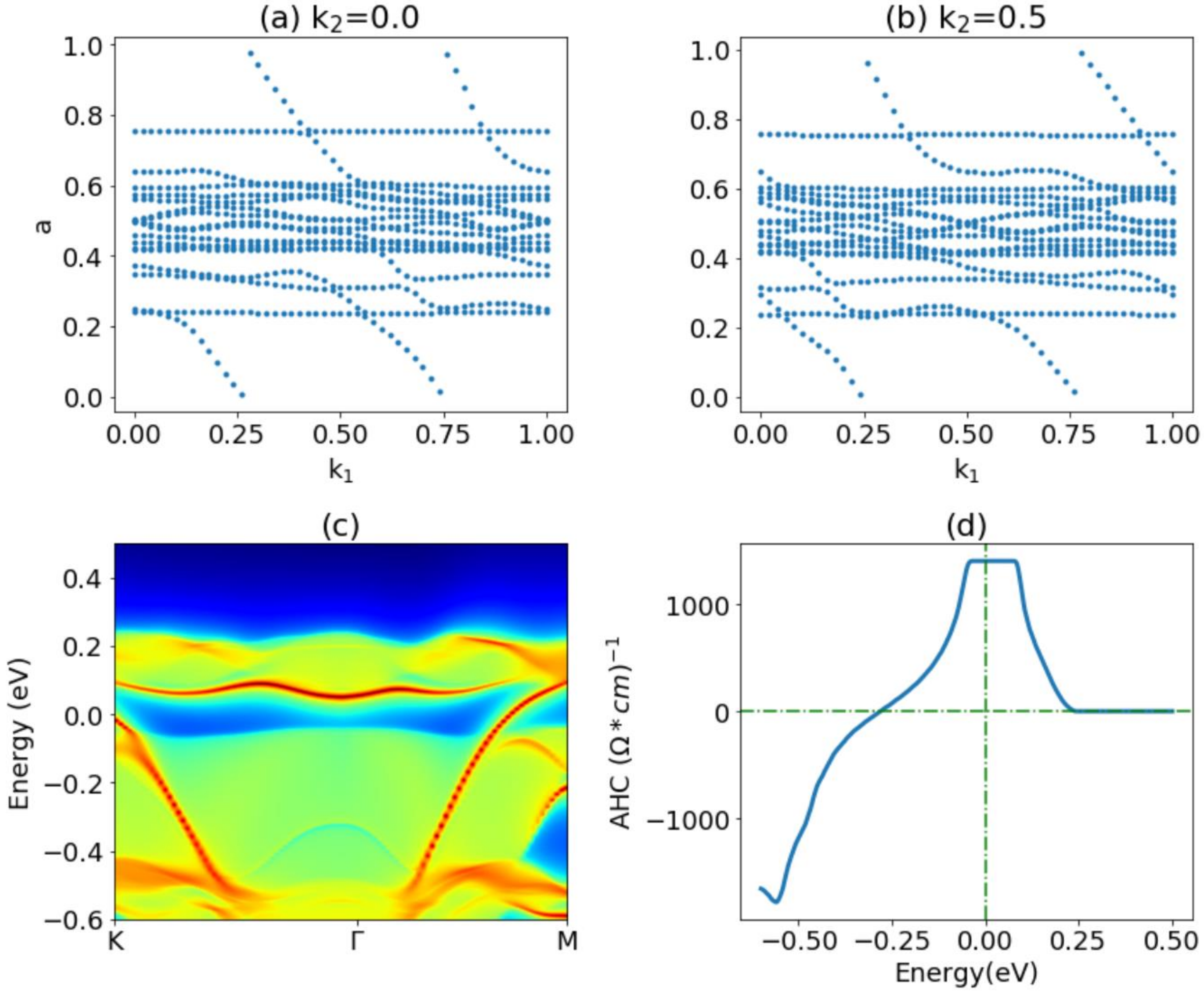

*Fig. 3 Wannier-charge center, surface bandstructure and anomalous hall conductivity for $NaRuO_2$ (JVASP-8122) with PBE+SOC. a) Wannier charge centers (WCC) for $k_1$=0.0, b) $k_1$=0.5, c) (001) surface bandstructure, d) AHC plot for the compound.*

In Fig. 4 we show the analysis of an example candidate topological metal $Y_3Sn$. $Y_3Sn$ crystallizes in $P6_3mmc$ space group and hexagonal system, has negative formation energy (-0.43 eV/atom) and 0.1 eV/atom energy above convex hull, suggesting that it should be experimentally synthesizable. The bandstructures in Fig. 1s show multiple band crossings for this system and has a spillage of 0.25. We plot the Fermi surface of this system in Fig. 4a which shows several conducting Fermi-channels represented by deep blue spots. The lighter colors indicate that there are not bands at the Fermi level. This material belongs to the Kagome lattice and such Fermi-surfaces have recently gained interest due to unique nodal line like features [44,45]. The (001) surface for this material also shows multiple bands crossing Fermi-level, which is shown in Fig. 4c. We observe several nodes in this material as shown in Fig. 4c with color coded energy level values. Energy levels with null value or blue color represents bands at Fermi level. The calculated anomalous Hall conductivity of this system is shown in Fig. 4d. The AHC is not quantized such as $NaRuO_2$, but still has a non-zero value at zero field which can be due to the topological features of the bandstructure. The SCAN+SOC and PBE+SOC bandstructure comparison for this system is also shown in the supplementary section (Fig. S2), which shows shifts in energy for several bands.

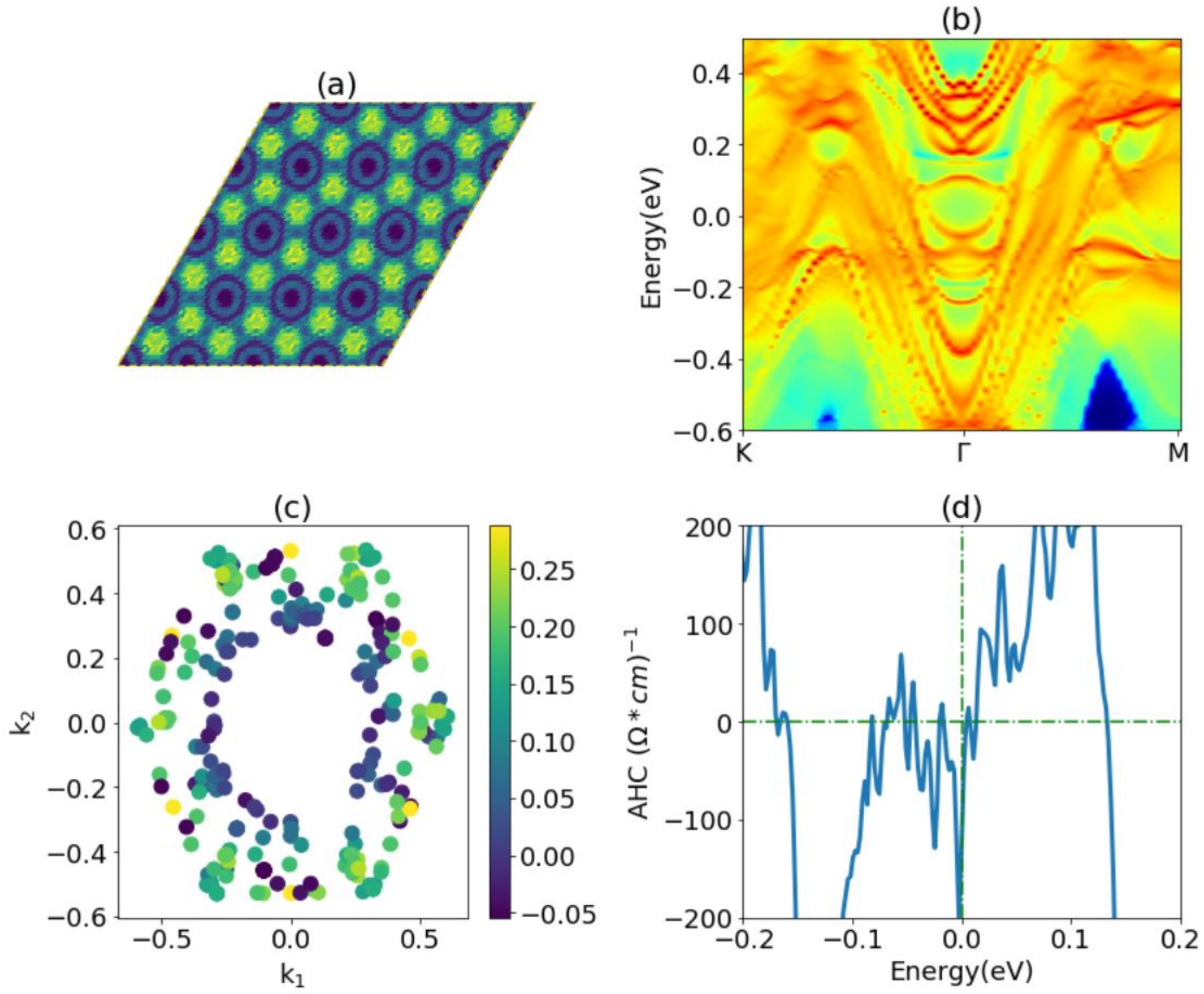

*Fig. 4 Analysis for $Y_3Sn$ (JVASP-37701) as a candidate semi-metal with PBE+SOC. a) Fermi-surface, b) (001) surface bandstructure, c) nodal points/lines, d) anomalous Hall conductivity.*

Next, in Fig. 5a, we show the likelihood that a compound containing a given element has a high-spillage for the 4734 materials screened from step a. More specifically, for every compound containing a given element, we calculate the percentage that have a spillage greater than 0.25. Consistent with known TMs, we observe that materials containing the elements such as Mn, Re, Fe, Ir, Pt, Bi and Pb are by far the likeliest ones to have high spillage. To contribute to SOC-induced band inversion, an element must both have significant SOC and contribute to bands located near the Fermi level, which favors heavy elements with moderate electronegativity. We use similar analysis for materials for thermoelectrics, solar cells, elastic constants etc. We can see some basic trends in the data but we intend to move towards more machine-learning prediction based on ML. To further accelerate the screening of magnetic topological materials we train three classification models using classical force-field inspired descriptors (CFID)[35] descriptors to predict the spillage, magnetic moment and bandgaps, based on data from the JARVIS-DFT database. The CFID descriptors provide a complete set of structural chemical features (1557 for each material) which we use with the Gradient Boosting Decision Tree (GBDT) algorithm as implemented in LightGBM[46] to train high accuracy ML models. The accuracy of the classification can be measured in terms of Receiver Operating Characteristic (ROC) Area Under Curve (AUC), which is 0.81 for spillage, and 0.97 for both the magnetic and bandgap models (using a 90 % to 10 % train test strategy). The ROC AUC is 0.5 for a random model, and 1.0 for a perfect model. The models trained for this work have ROC AUC greater than 0.81, signifying useful predictive power. The gradient boosting algorithm allows for feature importance to be extracted after training the model. Some of the high-importance descriptors of the ML models are: unfilled *d*-orbitals, and electronegativity which is intuitively reasonable. After training the ML models, we apply them on 1399770 materials from JARVIS, AFLOW[47], Materials-Project (MP)[48] and Open Quantum Materials Database (OQMD)[49] to find 77210 likely high-spillage materials using machine learning. The ML screened materials can then be subjected to the DFT workflow used in this work (see Fig. 1a) to further accelerate the search for magnetic topological materials. The ML models are distributed through the JARVIS-ML webapp.

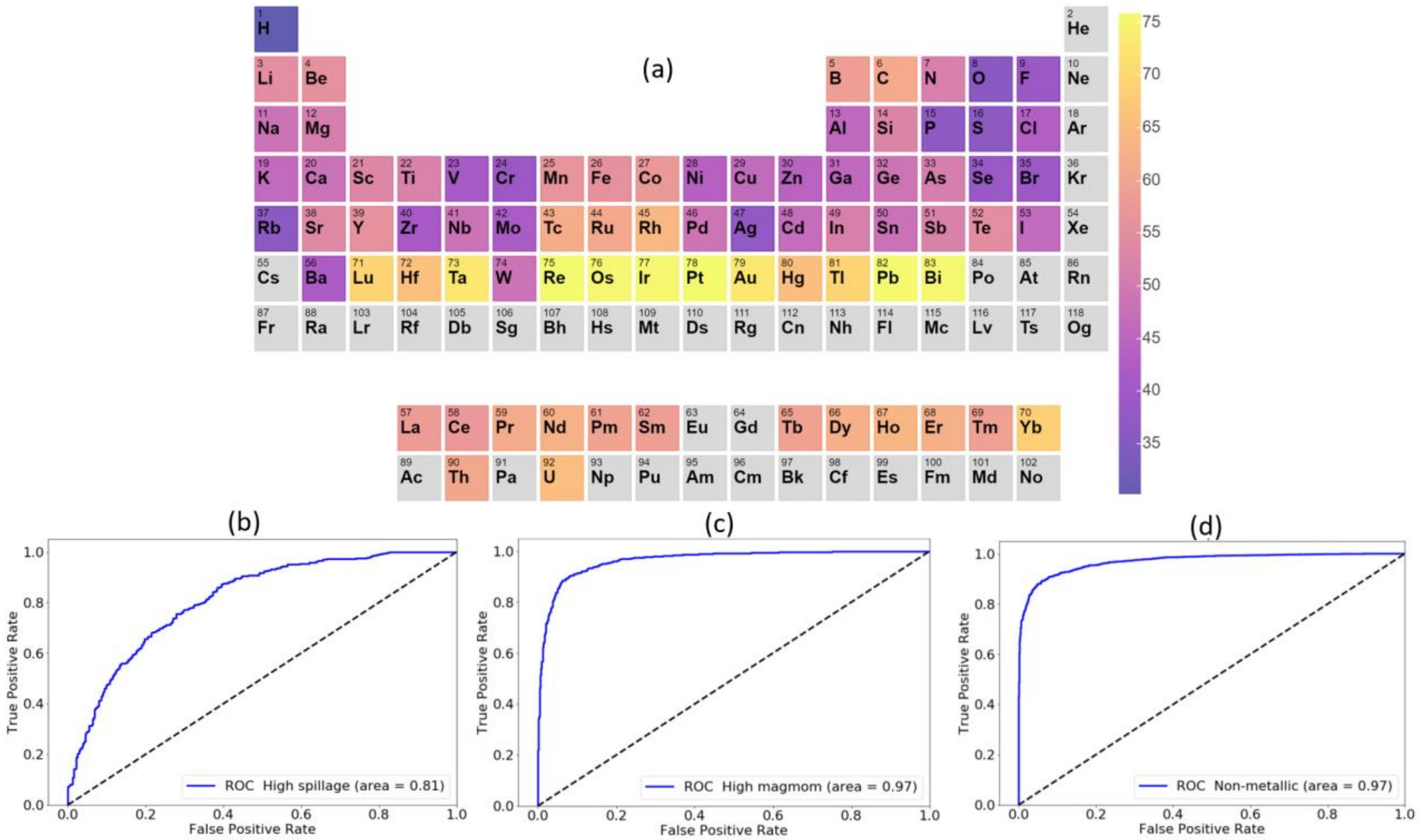

*Fig. 5 Periodic table trends and classification model receiver operating characteristics (ROC) curves. a) periodic table trends of compounds with high-spillage values. The elements in a material are weighed 1 or 0 if the material has high or low-values. Then the percentage probability of finding the element in a high-value material is calculated. b) For high/low spillage model (threshold 0.25), c) high/low magnetic moment (threshold 0.5 μB), d) Metals/non-metals based on electronic bandgaps (threshold 0.05 eV).*

Next, we discuss experimental results that support some of our theoretical findings. The AHE was first observed in ferromagnets where its origin lies in the interplay between spin–orbit coupling (SOC) and magnetization. Berry phase calculations have been proven accurate to predict SOC-induced intrinsic AHE in ferromagnets including Weyl (semi)metals, non-collinear antiferromagnets, non-coplanar magnets, and other nontrivial spin textures. In Fig. 6a, we show the experimental anomalous Hall conductivity as a function of magnetic field at 23 K, 25 K and 23 K for $CoNb_3S_6$. A large anomalous Hall conductivity at 23 K takes the value 27 $\Omega^{-1}$ $cm^{-1}$, which is a signature of experimental non-trivial band topology. Corresponding computational non-SOC, SOC bandstructures for this system, which has a maximum spillage value of 0.5 are shown in Fig. 2t. In Fig. 6b, we show the spin-pumping ferromagnetic resonance (SP-FMR) measurements by utilizing the inverse spin Hall effect (ISHE). In ISHE, a pure spin current $\vec{J}_S$ gets converted to a charge current $\vec{J}_C$ due to spin dependent asymmetric scattering phenomena. For spin pumping FMR measurements, $Mn_3Ge$ (100 nm)/ Permalloy (Py) (10 nm), Pt (10 nm)/Py (10 nm) and Py (10 nm) samples were prepared on sapphire substrate. A Pt device was also fabricated and analyzed because it provides an ideal benchmark for ISHE comparison. Fig. 6b shows the comparison between the ISHE charge current ($V_{ISHE}/R_{eq}$) for all three devices, where $R_{eq}$ is the total device resistance across the contact pads. Resistance values $R_{eq}$ for all devices were measured at room temperature in four-probe configuration. As expected, the Py single layer device is unaffected by ISHE, and thus $V_{sp}$ is entirely antisymmetric. On the other hand, the peak $V_{ISHE}/R_{eq}$ value of the $Mn_3Ge$/Py device is significantly larger than that of the Pt/Py device. The ratio of spin-Hall angles $\theta_{SH}^{Mn3Ge}/\theta_{SH}^{Pt}$ is estimated to be around 8 ± 2. The larger spin-Hall angle of $Mn_3Ge$ is a result of non-trivial band-topology which is consistent with the spillage signature.

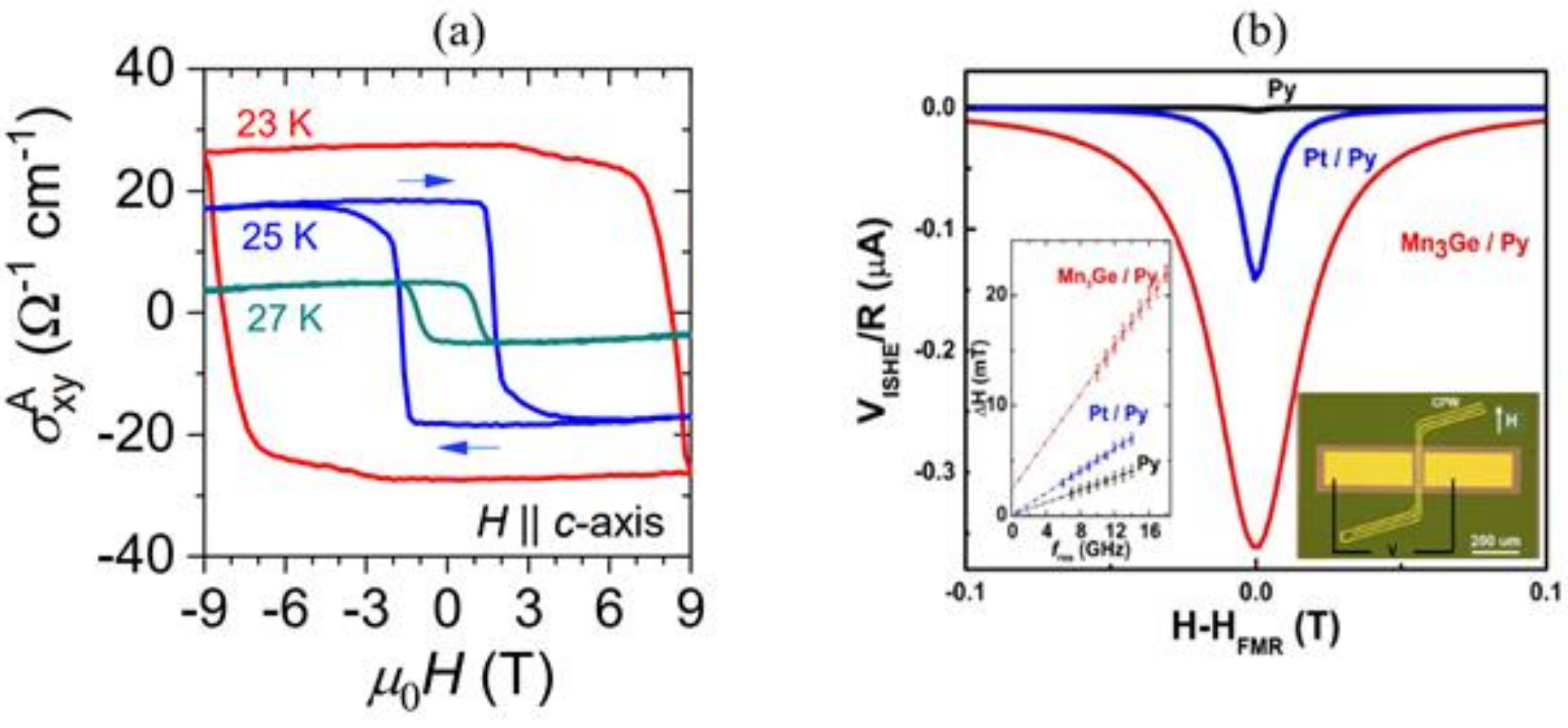

*Fig. 6 Experimental measurements of some of the candidate materials. a) anomalous Hall effect of $CoNb_3S_6$, b) comparison of inverse spin-Hall signal (symmetric component) among measured devices for $Mn_3Ge$. Right Inset: Optical image of the spin-pumping FMR device. Left Inset: Linear fit to the resonance linewidth ($\Delta H$) at various resonance frequencies.*

In summary, we have demonstrated the applicability of spin-orbit spillage, machine learning and experimental techniques to identify and characterize magnetic topological materials. We have also shown several remarkable trends in the topological chemistry with statistical analysis and periodic table distribution plots. Because we employ a high-throughput approach to screen a large database, we employ several assumptions, including assuming a ferromagnetic spin ordering and not performing detailed analysis of the dynamic or thermodynamic stability of our candidate materials. Detailed investigation of each material is out of the scope of this paper and will be undertaken in future work. We have made our datasets and tools publicly available to enhance the reproducibility and transparency of our work. We believe that our work can be of great help to guide future computational or experimental efforts to discover and characterize new magnetic topological materials.

**Methods**

**Density functional theory:** DFT calculations were carried out using the Vienna Ab-initio simulation package (VASP)[50,51] software using the workflow[59] given on our Github page (https://github.com/usnistgov/jarvis). We use the OptB88vdW functional[32], which gives accurate lattice parameters for both vdW and non-vdW (3D-bulk) solids[52]. We optimize the crystal-structures of the bulk and monolayer phases using VASP with OptB88vdW. The initial screening step for <1.5 eV bandgap materials is done with OptB88vdW bandgaps from the JARVIS-DFT database. Because SOC is not currently implemented for OptB88vdW in VASP, we carry out spin-polarized PBE and spin-orbit PBE calculations in order to calculate the spillage for each material. Such an approach has been validated by Refs. [23,53]. The crystal structure was optimized until the forces on the ions were less than 0.01 eV/Å and energy less than $10^{-6}$ eV. We use Wannier90[54] and Wannier-tools[55] to perform the Wannier-based evaluation of topological invariants.

As introduced in Ref.[56], we calculate the spin-orbit spillage, $\eta(\mathbf{k})$, given by the following equation:

$$\eta(\mathbf{k}) = n_{occ}(\mathbf{k}) - \mathrm{Tr}\big(P\tilde{P}\big) \tag{1}$$

where,

$P(\mathbf{k}) = \sum_{n=1}^{n_{occ}(\mathbf{k})} |\psi_{n\mathbf{k}}\rangle\langle\psi_{n\mathbf{k}}|$ is the projector onto the occupied wavefunctions without SOC, and $\tilde{P}$ is the same projector with SOC for band $n$ and k-point $\boldsymbol{k}$. We use a $\boldsymbol{k}$-dependent occupancy $n_{occ}(\mathbf{k})$ of the non-spin-orbit calculation so that we can treat metals, which have varying number of occupied electrons at each k-point[23]. Here, 'Tr' denotes trace over the occupied bands. We can write the spillage equivalently as:

$$\eta(\mathbf{k}) = n_{occ}(\mathbf{k}) - \sum_{m,n=1}^{n_{occ}(\mathbf{k})} |M_{mn}(\mathbf{k})|^2 \tag{2}$$

where $M_{mn}(\mathbf{k}) = \langle\psi_{m\mathbf{k}}|\tilde{\psi}_{n\mathbf{k}}\rangle$ is the overlap between occupied Bloch functions with and without SOC at the same wave vector $\boldsymbol{k}$. If the SOC does not change the character of the occupied wavefunctions, the spillage will be near zero, while band inversion will result in a large spillage. After spillage calculations, we run Wannier based Chern and $Z_2$-index calculations for these materials.

The Chern number, $C$ is calculated over the Brillouin zone, BZ, as:

$$C = \frac{1}{2\pi}\sum_n \int d^2\boldsymbol{k}\Omega_n \quad (3)$$

$$\Omega_n(\boldsymbol{k}) = -\mathrm{Im}\langle\nabla_{\boldsymbol{k}}u_{n\boldsymbol{k}}| \times |\nabla_{\boldsymbol{k}}u_{n\boldsymbol{k}}\rangle = \sum_{m\neq n}\frac{2\mathrm{Im}\langle\psi_{nk}|\hat{v}_x|\psi_{mk}\rangle\langle\psi_{mk}|\hat{v}_y|\psi_{nk}\rangle}{(\omega_m-\omega_n)^2} \quad (4)$$

Here, $\Omega_n$ is the Berry curvature, $u_{nk}$ being the periodic part of the Bloch wave in the $n$th band, $E_n = \hbar\omega_n$, $v_x$ and $v_y$ are velocity operators. The Berry curvature as a function of $\boldsymbol{k}$ is given by:

$$\Omega(\boldsymbol{k}) = \sum_n \int f_{nk}\Omega_n(\boldsymbol{k}) \quad (5)$$

Then, the intrinsic anomalous Hall conductivity (AHC) $\sigma_{xy}$ is given by:

$$\sigma_{xy} = -\frac{e^2}{\hbar}\int\frac{d^3\boldsymbol{k}}{(2\pi)^3}\Omega(\boldsymbol{k}) \quad (6)$$

In addition to searching for gapped phases, we also search for Dirac and Weyl semimetals by numerically searching for band crossings between the highest occupied and lowest unoccupied band, using the algorithm from WannierTools[55]. This search for crossings can be performed efficiently because it takes advantage of Wannier-based band interpolation. In an ideal case, the band crossings will be the only points at the Fermi level; however, in most cases, we find additional trivial metallic states at the Fermi level. The surface spectrum was calculated by using the Wannier functions and the iterative Green's function method[57,58] .

Starting from ~40000 materials in the JARVIS-DFT database, we screened for materials with magnetic moment >0.5 μB and having heavy elements (atomic weight ≥ 65) and bandgaps <1.5 eV. After carrying out spin-orbit spillage calculations on them, we broadly classify them into insulators and semimetals with non-vanishing and vanishing electronic bandgaps. For materials with high spillage, we run Wannier calculations to calculate the Chern number, anomalous hall conductivity, surface bandstructures and Fermi-surfaces. We also run SCAN functional based calculations on the high spillage materials to check the changes in bandgaps and magnetic moments. So far, we have calculated 11483 SOSs for both magnetic/non-magnetic, metallic/non-metallic systems.

**Machine learning model:**

The machine-learning models are trained using classical force-field inspired descriptors (CFID) descriptors and supervise machine learning techniques using gradient boosting techniques in the LightGBM[46] package[59]. The CFID gives a unique representation of a material using structural (such as radial, angle and dihedral distributions), chemical, and charge descriptors. The CFID provides 1557 descriptors for each material. We use 'VarianceThreshold' and 'StandardScaler' preprocessing techniques available in scikit-learn before applying the ML technique to remove low-variance descriptors and standardize the descriptor set. We use DFT data for developing machine learning models for high/low spillage (threshold 0.5), high/low magnetic moment (threshold 0.5 μB), high/low bandgap (threshold 0.0 eV) to further accelerate the screening process

The CFID has been recently used to develop several high-accuracy ML models for material properties such as formation energies, bandgaps, refractive index, bulk and shear modulus and exfoliation energies k-points, cut-offs, and solar-cell efficiencies. The accuracy of the model is evaluated based on area under curve (AUC) for the receiver operating characteristic (ROC). We provide a sample script for the ML training in the supplementary information.

**Experimental details:**

**$CoNb_3S_6$**:

Single crystals of $CoNb_3S_6$ were grown by chemical vapor transport using iodine as the transport agent[59]. First, a polycrystalline sample was prepared by heating stoichiometric amounts of cobalt powder (Alfa Aesar 99.998 %), niobium powder (Johnson Matthey Electronics 99.8 %), and sulfur pieces (Alfa Aesar 99.9995 %) in an evacuated silica ampoule at 900 °C for 5 days. Subsequently, 2 g of the powder was loaded together with 0.5 g of iodine in a fused silica tube of 14 mm inner diameter. The tube was evacuated and sealed under vacuum. The ampoule of 11 cm length was loaded in a horizontal tube furnace in which the temperature of the hot zone was kept at 950 °C and that of the cold zone was ≈850 °C for 7 days. Several $CoNb_3S_6$ crystals formed with a distinct, well-faceted flat plate-like morphology. The crystals of $CoNb_3S_6$ were examined by single crystal X-ray diffraction at room temperature. Compositional analysis was done using an energy dispersive X-ray spectroscopy (EDS) at the Electron Microscopy Center, ANL.

Transport measurements were performed on a quantum design PPMS following a conventional 4-probe method. Au wires of 25 μm diameter were attached to the sample with Epotek H20E silver epoxy. An electric current of 1 mA was used for the transport measurements. The following method was adopted for the contact misalignment correction in Hall effect measurements. The Hall resistance was measured at H = 0 by decreasing the field from the positive magnetic field (RH+), where H represents the external magnetic field. Again, the Hall resistance was measured at H = 0 by increasing the field from negative magnetic field (RH−). Average of the absolute value of (RH+) and (RH−) was then subtracted from the measured Hall resistance. The conventional antisymmetrization method was also used for the Hall resistance measured at 28 K (above TN) and at 2 K (where no anomalous Hall effect was observed), which gave same result as obtained from the former method.

**$Mn_3Ge$:**

In ISHE, a pure spin current $\vec{J}_S$ gets converted to a charge current $\vec{J}_C$ due to spin dependent asymmetric scattering phenomena[59]. To maximize the ISHE signal, the external magnetic field is applied along $[1\bar{1}00]$ and dc voltage is measured along $[11\bar{2}0]$ directions. An optical image of the spin-pumping device is shown in Fig. 6b. For spin pumping FMR measurements, (i) $Mn_3Ge$ (100 nm)/ Py (10 nm) and (ii) Pt (10 nm) Py (10 nm) (iii) Py (10 nm) samples were prepared on sapphire substrate. They were fabricated into 1000 μm×200 μm bars by photolithography and ion milling. Coplanar waveguides (CPW) with 170-nm thick Ti (20 nm)/Au (150 nm) were subsequently fabricated. Using ICP-CVD method, an additional SiN (150 nm) layer is deposited between CPW and the sample for electric isolation. The microwave frequencies were tuned between 10 GHz to 18 GHz with varying power (12 dBm - 18 dBm) while magnetic field was swept between -0.4 T to 0.4 T along the CPW axis. Measurements were performed at room temperature and field resolution of 2 mT was adopted throughout.

**Data availability**

JARVIS-related data is available at the JARVIS-API (http://jarvis.nist.gov), and JARVIS-DFT (https://jarvis.nist.gov/jarvisdft/) webpages.

## Code availability

Python-language based codes with examples are available at JARVIS-tools page: https://github.com/usnistgov/jarvis.

## Contributions

K.C. designed the computational workflows, carried out high-throughput calculations, analysis, and developed the websites. K.F.G helped in developing the workflow and analysis of the data. N.J.G. performed the experiments for $CoNb_3S_6$. N.A. performed the experiments for $Mn_3Ge$. All authors contributed to writing the manuscript.

## Acknowledgements

K.C., K.F.G., and F.T. thank the National Institute of Standards and Technology for funding, computational, and data-management resources. NJG acknowledges support from U.S. Department of Energy, Office of Science, Basic Energy Sciences, Materials Science and Engineering Division.

# Supplementary information: High-throughput search for magnetic topological materials using spin-orbit spillage, machine-learning and experiments

Kamal Choudhary[1,2], Kevin F. Garrity[1], Nirmal J. Ghimire[3,4], Naween Anand[5], Francesca Tavazza[1]

1 Materials Science and Engineering Division, National Institute of Standards and Technology, Gaithersburg, MD, 20899, USA.

2 Theiss Research, La Jolla, CA 92037, USA.

3. Department of Physics and Astronomy, George Mason University, Fairfax, VA 22030, USA.

4. Quantum Science and Engineering Center, George Mason University, Fairfax, VA 22030, USA.

5. Materials Science Division, Argonne National Laboratory, Argonne, Illinois 60439, USA.

Table S1 Bandgap comparison for PBE+SOC and SCAN+SOC functionals with 1000/atom k-point settings.

| JID | PBE+SOC(eV) | SCAN+SOC(eV) |
|---|---|---|
| **JVASP-44705** | 0.02 | 1.2337 |
| **JVASP-2817** | 0.37 | 1.3048 |
| **JVASP-12648** | 0.27 | 1.1202 |
| **JVASP-10484** | 0.13 | 0.9678 |
| **JVASP-57388** | 0.14 | 0.91 |
| **JVASP-16408** | 0.28 | 0.5995 |
| **JVASP-55805** | 0.09 | 0.35 |
| **JVASP-21389** | 0.21 | 0.4312 |
| **JVASP-42538** | 0.03 | 0.2225 |
| **JVASP-8385** | 0.2 | 0.3273 |
| **JVASP-44991** | 0.04 | 0.116 |
| **JVASP-49890** | 0.01 | 0.072 |
| **JVASP-43466** | 0.03 | 0.0865 |
| **JVASP-82132** | 0.03 | 0.0528 |
| **JVASP-26528** | 0.35 | 0.3722 |
| **JVASP-81597** | 0.53 | 0.5518 |
| **JVASP-8384** | 0.01 | 0.0266 |
| **JVASP-16409** | 0.3 | 0.3104 |
| **JVASP-52135** | 0 | 0.0046 |
| **JVASP-59757** | 0 | 0 |
| **JVASP-17989** | 0 | 0 |
| **JVASP-43095** | 0 | 0 |
| **JVASP-17898** | 0 | 0 |
| **JVASP-16012** | 0 | 0 |
| **JVASP-59630** | 0 | 0 |
| **JVASP-55644** | 0 | 0 |
| **JVASP-8538** | 0 | 0 |
| **JVASP-59509** | 0 | 0 |
| **JVASP-26231** | 0 | 0 |
| **JVASP-39287** | 0 | 0 |
| **JVASP-26527** | 0 | 0 |
| **JVASP-38244** | 0 | 0 |
| **JVASP-38246** | 0 | 0 |
| **JVASP-38201** | 0 | 0 |
| **JVASP-38157** | 0 | 0 |
| **JVASP-38248** | 0 | 0 |
| **JVASP-17268** | 0 | 0 |
| **JVASP-37701** | 0 | 0 |
| **JVASP-45925** | 0 | 0 |
| **JVASP-34486** | 0 | 0 |
| **JVASP-81304** | 0 | 0 |
| **JVASP-76959** | 0 | 0 |
| **JVASP-80606** | 0 | 0 |
| **JVASP-81275** | 0 | 0 |
| **JVASP-81033** | 0.01 | 0 |
| **JVASP-76869** | 0.06 | 0.048 |
| **JVASP-80151** | 0.03 | 0.0142 |
| **JVASP-24841** | 0.04 | 0.0205 |
| **JVASP-16201** | 0.02 | 0 |
| **JVASP-18368** | 0.02 | 0 |
| **JVASP-44742** | 0.02 | 0 |
| **JVASP-76813** | 0.02 | 0 |
| **JVASP-77062** | 0.17 | 0.1493 |
| **JVASP-50689** | 0.03 | 0 |
| **JVASP-17265** | 0.39 | 0.3555 |
| **JVASP-79685** | 0.29 | 0.2462 |
| **JVASP-22442** | 0.16 | 0.1138 |
| **JVASP-81240** | 0.15 | 0.103 |
| **JVASP-8122** | 0.11 | 0.056 |
| **JVASP-21125** | 0.2 | 0.1449 |
| **JVASP-76930** | 0.16 | 0.1028 |
| **JVASP-21502** | 0.06 | 0 |
| **JVASP-26681** | 0.18 | 0.1136 |
| **JVASP-21417** | 0.07 | 0 |

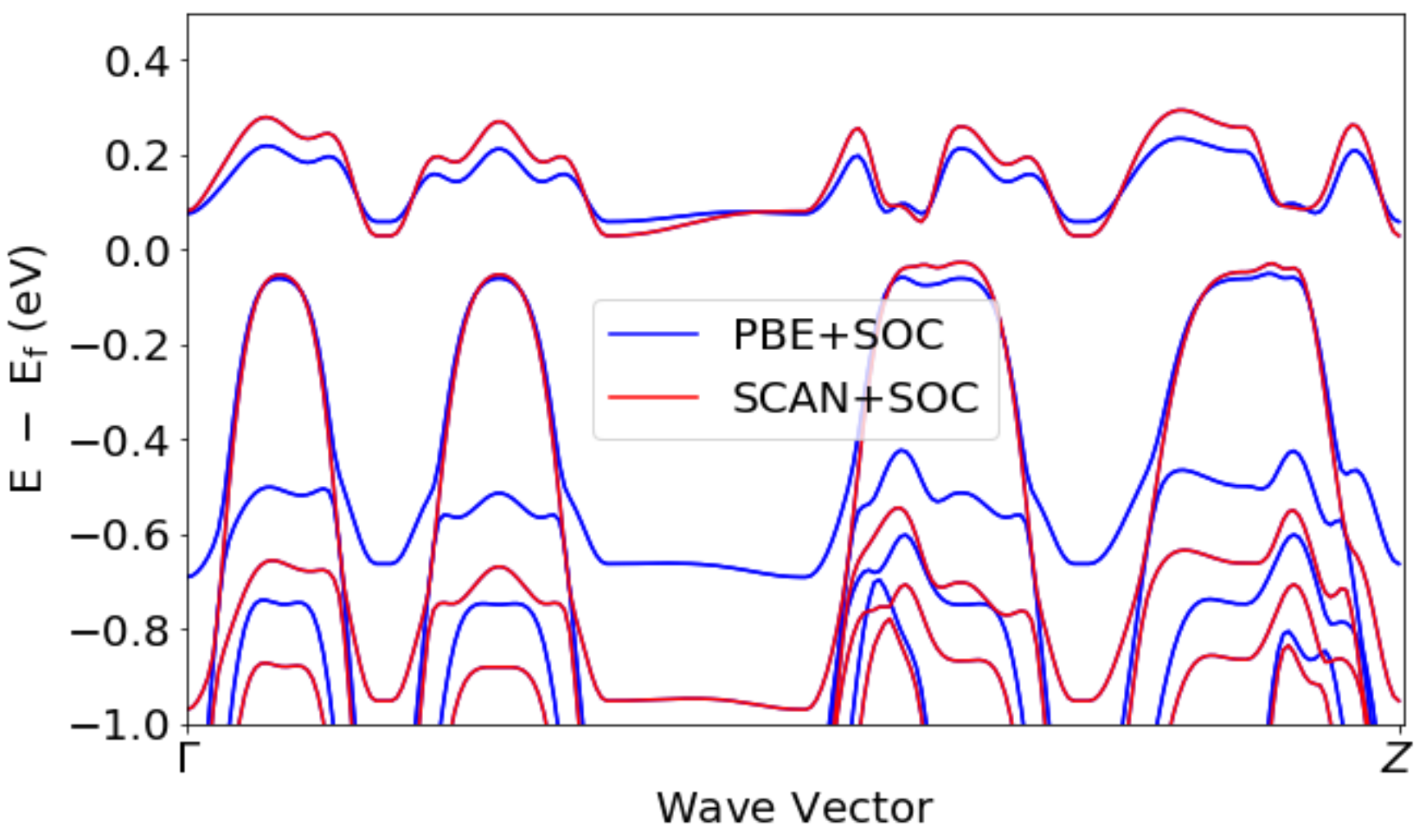

Fig. S1 PBE+SOC and SCAN+SOC bandstructures for $NaRuO_2$.

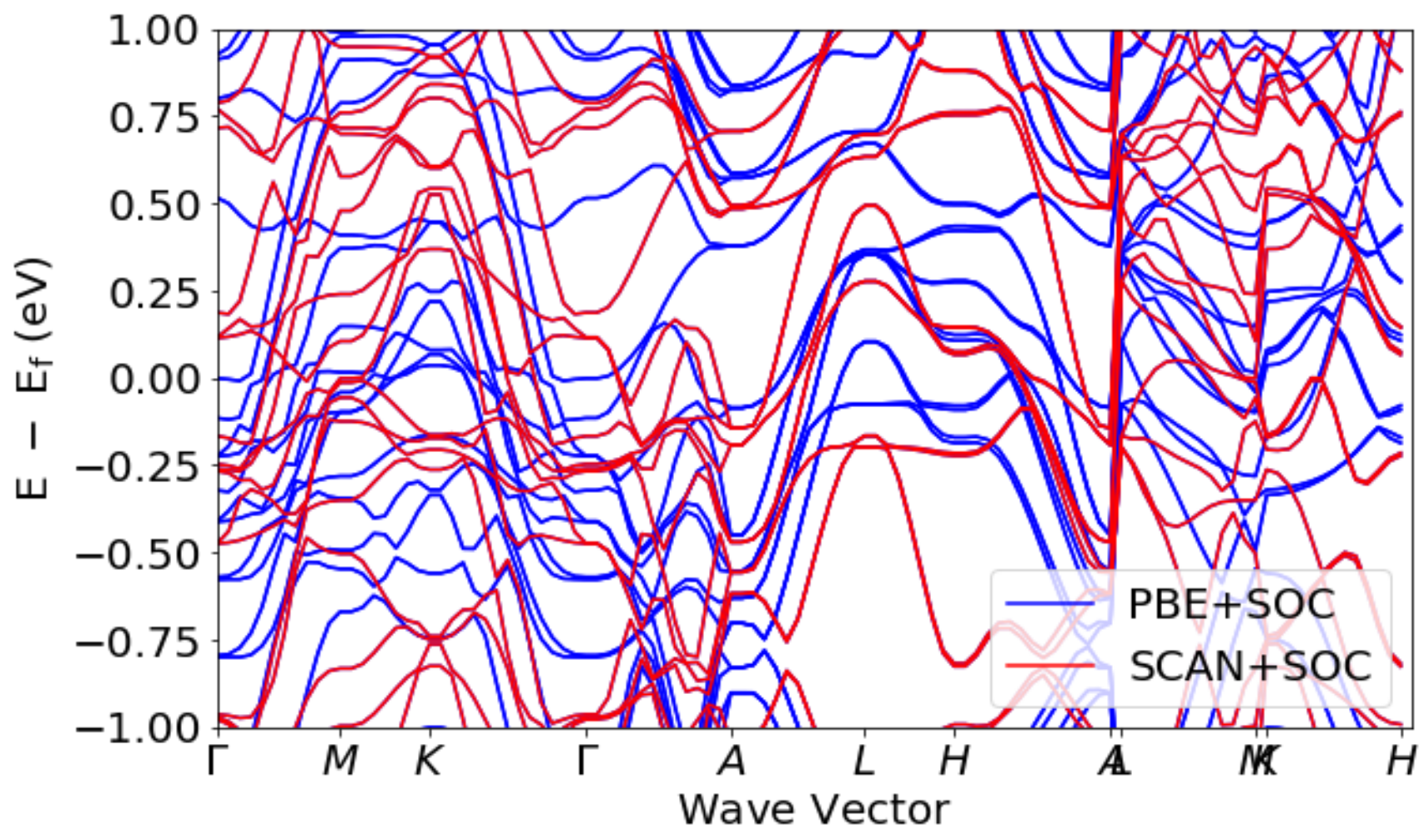

Fig. S2 PBE+SOC and SCAN+SOC bandstructures for $Y_3Sn$.

Table S2 High-spillage material candidates.

| JID | Formula | Spillage | Gap | Magmom | Ehull |
|---|---|---|---|---|---|
| **JVASP-53555** | MnIn2Te4 | 0.828 | 0 | 4.963 | 0 |
| **JVASP-53226** | Ca5MnPb3 | 7.203 | 0 | 9.001 | 0 |
| **JVASP-15074** | YCo5 | 0.376 | 0 | 6.796 | 0 |
| **JVASP-14699** | MnTe | 1.038 | 0 | 8.073 | 0.012265 |
| **JVASP-60253** | UTe3 | 3.323 | 0 | 4.89 | 0 |
| **JVASP-84812** | RbMnAs | 0.44 | 0 | 3.772 | 0 |
| **JVASP-53303** | MnTe | 0.293 | 0 | 13.39 | 0.061846 |
| **JVASP-43433** | Li2NbFe3O8 | 2.936 | 0 | 3 | 0 |
| **JVASP-60477** | RuCl3 | 0.264 | 0 | 1.985 | 0 |
| **JVASP-43643** | Li2CoSnO4 | 0.322 | 0 | 2.789 | 0.003284 |
| **JVASP-43846** | MnP2WO8 | 1.539 | 0 | 6 | 0 |
| **JVASP-16176** | Li2NbF6 | 0.432 | 0 | 1 | 0 |
| **JVASP-18583** | TiAgHg2 | 1.104 | 0 | 1.649 | 0.234241 |
| **JVASP-18620** | Mn2Au5 | 0.282 | 0 | 8.357 | 0.0225 |
| **JVASP-48582** | Li6Mn5SbO2 | 5.953 | 0 | 21.945 | 0 |
| **JVASP-49957** | Li2RhO3 | 0.459 | 0 | 1.826 | 0 |
| **JVASP-14962** | PuFe2 | 4.719 | 0 | 3.591 | 0 |
| **JVASP-84911** | RbMnAs | 0.367 | 0 | 3.967 | 0 |
| **JVASP-18519** | SrMnBi2 | 1.428 | 0 | 7.883 | 0 |
| **JVASP-18467** | K2MoCl6 | 0.263 | 0 | 2 | 0 |
| **JVASP-18551** | RbMnSe2 | 0.302 | 0 | 4 | 0 |
| **JVASP-18468** | K2OsBr6 | 0.934 | 0 | 1.994 | 0 |
| **JVASP-14964** | Mn3Rh | 0.815 | 0 | 0.903 | 0.059416 |
| **JVASP-14584** | CrPt3 | 2.521 | 0 | 2.791 | 0 |
| **JVASP-15006** | CrSb2 | 0.545 | 0 | 4.192 | 0.205306 |
| **JVASP-44114** | Li2V3SnO8 | 1.008 | 0 | 0.988 | 0.038036 |
| **JVASP-43542** | Li3YNi2O6 | 2.944 | 0 | 1.998 | 0 |
| **JVASP-15562** | RbFeS2 | 0.278 | 0 | 3.13 | 0 |
| **JVASP-48117** | Li2Fe3TeO8 | 3.823 | 0 | 2.025 | 0.003622 |
| **JVASP-60294** | UIN | 1.201 | 0.03 | 4 | 0 |
| **JVASP-43107** | Li2Fe3SnO8 | 0.252 | 0 | 4 | 0.031039 |
| **JVASP-14926** | VSb | 1.02 | 0 | 2.891 | 0.123297 |
| **JVASP-20094** | Co3Mo | 1.016 | 0 | 0.9 | 0 |
| **JVASP-54386** | NbFeO4 | 0.523 | 0 | 2.015 | 0 |
| **JVASP-18625** | VAu2 | 2.102 | 0 | 3.782 | 0.022372 |
| **JVASP-15649** | MnCuSb | 1.015 | 0 | 4.296 | 0.187604 |
| **JVASP-14711** | FeAs | 0.919 | 0 | 1.923 | 0.002932 |
| **JVASP-14928** | MnNbSi | 1.02 | 0 | 2.14 | 0 |
| **JVASP-11974** | NbCo3 | 1.094 | 0 | 11.578 | 0.078038 |
| **JVASP-87133** | UTe2 | 2.541 | 0 | 4.916 | 0 |
| **JVASP-43362** | Li2Fe3TeO8 | 0.469 | 0 | 9.606 | 0 |
| **JVASP-15567** | MnCo2Sb | 0.304 | 0 | 5.793 | 0.041562 |
| **JVASP-43661** | Li4Mn3Co3Sn2O6 | 4.974 | 0 | 8.016 | 0 |
| **JVASP-15104** | Mn3ZnC | 0.551 | 0 | 6.801 | 0.009448 |
| **JVASP-15237** | HfGaCo2 | 4.98 | 0 | 0.876 | 0 |
| **JVASP-42681** | Ti2Co3Te3O6 | 3.082 | 0 | 9.495 | 0 |
| **JVASP-19910** | CoPt | 1.378 | 0 | 4.447 | 0 |
| **JVASP-14889** | RuF3 | 1.989 | 0 | 1.975 | 0 |
| **JVASP-50360** | Bi2PO6 | 0.329 | 0 | 0.737 | 0 |
| **JVASP-52375** | HfMn6Sn6 | 0.391 | 0 | 13.037 | 0.009234 |
| **JVASP-54439** | FeB2W2 | 0.645 | 0 | 3.35 | 0 |
| **JVASP-43466** | Li2Fe3SbO8 | 5.063 | 0.03 | 3 | 0.012412 |
| **JVASP-18924** | FePt3 | 1.348 | 0 | 4.197 | 0 |
| **JVASP-21561** | Nb4CrS8 | 1.024 | 0 | 4.931 | 0 |
| **JVASP-14934** | CrCuSe2 | 0.296 | 0 | 3 | 0.254504 |
| **JVASP-43125** | Cr3SbP4O6 | 3.53 | 0.12 | 9 | 0 |
| **JVASP-42156** | Li4Fe3SbO8 | 2.151 | 0 | 8.995 | 0 |
| **JVASP-48191** | Li2P2WO8 | 1.162 | 0 | 4 | 0.007233 |
| **JVASP-50895** | PrTlO3 | 0.721 | 0 | 2 | 0 |
| **JVASP-18525** | Rb2WBr6 | 0.569 | 0 | 2 | 0 |
| **JVASP-52305** | Ga2CuO4 | 0.393 | 0 | 1.989 | 0 |
| **JVASP-16908** | TiBi2O6 | 1.081 | 0 | 1.891 | 0.252842 |
| **JVASP-16201** | MgRhF6 | 0.902 | 0.02 | 0.991 | 0 |
| **JVASP-53381** | Cr4Cu3Se8 | 0.629 | 0 | 11.097 | 0.037619 |
| **JVASP-53248** | Co2B6W2 | 0.537 | 0 | 17.019 | 0 |
| **JVASP-47467** | TaFeO4 | 0.69 | 0 | 2 | 0.016797 |
| **JVASP-19524** | UI3 | 2.823 | 0.04 | 6 | 0 |
| **JVASP-50977** | Li2CuBiO4 | 0.594 | 0 | 0.723 | 0 |
| **JVASP-15326** | CaMn2Sb2 | 0.517 | 0 | 7.131 | 0 |
| **JVASP-16161** | ScFeGe | 0.446 | 0 | 4.543 | 0 |
| **JVASP-42352** | HfCrO4 | 2.027 | 0 | 8 | 0 |
| **JVASP-15617** | SrCo2P2 | 0.266 | 0 | 0.591 | 0 |
| **JVASP-53515** | Mn2SbTe | 0.333 | 0 | 15.725 | 0.298565 |
| **JVASP-16001** | K2RhF6 | 1.338 | 0.13 | 0.994 | 0 |
| **JVASP-6145** | US3 | 1.769 | 0 | 4 | 0 |
| **JVASP-60459** | UTe3 | 1.707 | 0 | 3.409 | 0 |
| **JVASP-18528** | Tl2WCl6 | 1.043 | 0 | 2 | 0 |
| **JVASP-53265** | AlFe2Mo | 0.389 | 0 | 0.789 | 0 |
| **JVASP-50093** | TiNbO4 | 0.654 | 0 | 0.533 | 0 |
| **JVASP-43389** | Li4MnW3O2 | 1.119 | 0.42 | 4.99 | 0 |
| **JVASP-52989** | Sr3Mn4O2 | 0.462 | 0 | 9.994 | 0 |
| **JVASP-43935** | Li4Mn3V3Sn2O6 | 2.408 | 0 | 14.756 | 0 |
| **JVASP-43390** | Li4Cr5SbO2 | 7.516 | 0 | 3 | 0 |
| **JVASP-43694** | CdFeO3 | 0.395 | 0 | 12.87 | 0 |
| **JVASP-15511** | FeAgS2 | 0.276 | 0 | 2.056 | 0.054059 |
| **JVASP-726** | CrS2 | 0.308 | 0 | 2.047 | 0 |
| **JVASP-14679** | VPt3 | 1.294 | 0 | 1.357 | 0.014238 |
| **JVASP-19895** | CoPt3 | 1.687 | 0 | 2.833 | 0 |
| **JVASP-15663** | MnSbAu | 0.904 | 0 | 4.559 | 0.151837 |
| **JVASP-53544** | Cr4Cu3Te8 | 1.05 | 0 | 11.359 | 0 |
| **JVASP-47523** | Co3SbO8 | 0.28 | 0 | 3.696 | 0 |
| **JVASP-42538** | Ta2CrNO5 | 2.788 | 0.03 | 6 | 0 |
| **JVASP-6097** | VCl3 | 0.973 | 0 | 4 | 0 |
| **JVASP-50922** | CuAuO2 | 0.456 | 0 | 0.51 | 0.012338 |
| **JVASP-20076** | FeSe | 0.421 | 0 | 4.509 | 0.198073 |
| **JVASP-18532** | CaMn2Bi2 | 1.171 | 0 | 8.652 | 0.264587 |
| **JVASP-49890** | Ba4IrO6 | 2.44 | 0.01 | 1.651 | 0 |
| **JVASP-43616** | LiFeSbO4 | 0.354 | 0 | 8.13 | 0.052363 |
| **JVASP-51442** | GaFe2Co | 0.735 | 0 | 5.679 | 0.07515 |
| **JVASP-15583** | MnAlAu2 | 0.289 | 0 | 3.785 | 0 |
| **JVASP-20110** | MnPt3 | 2.412 | 0 | 4.245 | 0 |
| **JVASP-6766** | HfFeCl6 | 1.042 | 0.02 | 4 | 0 |
| **JVASP-15211** | FeSiRu2 | 0.394 | 0 | 3.686 | 0 |
| **JVASP-50689** | Li4Ti3Cu3Te2O6 | 1.026 | 0.03 | 2.771 | 0 |
| **JVASP-1867** | FeAgO2 | 0.75 | 0 | 1.032 | 0.000927 |
| **JVASP-16213** | MnNbGe | 0.381 | 0 | 3.711 | 0 |
| **JVASP-14419** | CoI2 | 0.852 | 0 | 2.409 | 0 |
| **JVASP-13600** | ZrFeCl6 | 1.015 | 0.03 | 4 | 0 |
| **JVASP-57304** | Sr3Fe2Cu2Se2O5 | 1.014 | 0 | 7.634 | 0 |
| **JVASP-19831** | MnSb | 1.005 | 0 | 6.273 | 0.0527 |
| **JVASP-56996** | TaFe2 | 2.098 | 0 | 2.382 | 0 |
| **JVASP-58164** | Mg2TiIrO6 | 2.476 | 0 | 1.21 | 0 |
| **JVASP-56764** | Tl4CrI6 | 0.295 | 0 | 8 | 0.00395 |
| **JVASP-58491** | V4ZnS8 | 1.181 | 0 | 1.181 | 0 |
| **JVASP-57228** | Sr2Mn3Bi2O2 | 0.516 | 0 | 11.681 | 0 |
| **JVASP-57886** | BaCr4O8 | 0.354 | 0 | 9.992 | 0 |
| **JVASP-56321** | RbTiBr3 | 0.347 | 0 | 2.621 | 0 |
| **JVASP-15423** | TlCrTe2 | 0.401 | 0 | 3.054 | 0 |
| **JVASP-49947** | Na2IrO3 | 1.294 | 0 | 1.798 | 0 |
| **JVASP-57036** | Sr2CoCl2O2 | 0.421 | 0 | 2.108 | 0 |
| **JVASP-58165** | Ca2TiIrO6 | 2.291 | 0 | 0.82 | 0 |
| **JVASP-57153** | FeBiO3 | 0.322 | 0 | 2.811 | 0.100464 |
| **JVASP-54622** | GaFe2Ni | 0.255 | 0 | 4.509 | 0.08455 |
| **JVASP-15424** | MnRh2Pb | 0.881 | 0 | 4.674 | 0 |
| **JVASP-51466** | NiRh2O4 | 0.359 | 0 | 3.986 | 8.42E-05 |
| **JVASP-58166** | TiZn2IrO6 | 1.96 | 0 | 1.126 | 0 |
| **JVASP-35684** | Mn3Sn | 0.477 | 0 | 1.01 | 0.216904 |
| **JVASP-58167** | Mg2SnIrO6 | 1.657 | 0 | 0.865 | 0 |
| **JVASP-21107** | Mn3P6Pd20 | 1.106 | 0 | 14.047 | 0 |
| **JVASP-57905** | BaTi4O7 | 0.408 | 0 | 3.473 | 0.119577 |
| **JVASP-15256** | FeCuPt2 | 1.662 | 0 | 3.93 | 0 |
| **JVASP-54890** | RbW3Cl9 | 0.579 | 0.06 | 1.616 | 0 |
| **JVASP-58177** | Mg2MnIrO6 | 1.523 | 0 | 4.194 | 0 |
| **JVASP-59902** | ZnCr2S4 | 0.303 | 0 | 12 | 0.158825 |
| **JVASP-58503** | ZnFe4S8 | 1.05 | 0 | 6.979 | 0 |
| **JVASP-35299** | MnCoSn | 1.013 | 0 | 8.783 | 0.348821 |
| **JVASP-59607** | YMnGe | 0.411 | 0 | 11.184 | 0 |
| **JVASP-58179** | Mg2CoIrO6 | 1.673 | 0 | 2.154 | 0 |
| **JVASP-15693** | Mn2Sb | 0.5 | 0 | 11.488 | 0.333028 |
| **JVASP-15730** | ThCo2Si2 | 1.149 | 0 | 0.97 | 0 |
| **JVASP-51048** | FeCu2SnS4 | 0.313 | 0 | 3.874 | 0.006128 |
| **JVASP-15428** | KCo2Se2 | 1.006 | 0 | 2.002 | 0 |
| **JVASP-56922** | MnAsRh | 0.347 | 0 | 9.719 | 0 |
| **JVASP-51894** | FeBiO3 | 2.269 | 0 | 13.908 | 0.005442 |
| **JVASP-57388** | MnBiAsO5 | 3.258 | 0.14 | 10 | 0 |
| **JVASP-57951** | YCu3Sn4O2 | 1.146 | 0 | 2.015 | 0 |
| **JVASP-58180** | Zn2CoIrO6 | 1.813 | 0 | 1.382 | 0 |
| **JVASP-54898** | SrCo2As2 | 0.471 | 0 | 0.846 | 0 |
| **JVASP-16042** | TaF3 | 0.497 | 0 | 1.99 | 0.448904 |
| **JVASP-18575** | Rb2WCl6 | 0.591 | 0 | 2 | 0 |
| **JVASP-58507** | Ca2CuIrO6 | 2.84 | 0 | 3.717 | 0 |
| **JVASP-49616** | ZnFeO3 | 0.9 | 0 | 8.717 | 0.063702 |
| **JVASP-57848** | Ba6Ru3Cl2O2 | 0.513 | 0 | 3.365 | 0 |
| **JVASP-55136** | Fe3PtN | 1.033 | 0 | 7.954 | 0 |
| **JVASP-57242** | Sr2Fe2S2OF2 | 0.472 | 0 | 7.526 | 0 |
| **JVASP-59737** | Ta2CuO6 | 0.354 | 0 | 1.997 | 0 |
| **JVASP-21401** | NaSr3IrO6 | 3.184 | 0.23 | 3.893 | 0 |
| **JVASP-54949** | Co2Sn | 1.118 | 0 | 2.772 | 0.219942 |
| **JVASP-19286** | Y2Fe2O7 | 3.858 | 0 | 8.012 | 0 |
| **JVASP-57319** | Ba2YFe3O8 | 0.506 | 0 | 4.796 | 0 |
| **JVASP-8616** | Sr2CoBr2O2 | 0.282 | 0 | 2.128 | 0 |
| **JVASP-55156** | Fe3RhN | 1.011 | 0 | 8.232 | 0 |
| **JVASP-59673** | Y2Ru2O7 | 2.262 | 0 | 1.468 | 0 |
| **JVASP-57430** | Ba2Mn2Bi2O | 2.292 | 0 | 19.266 | 0 |
| **JVASP-15076** | FePd | 0.28 | 0 | 6.495 | 0.024811 |
| **JVASP-15057** | Fe3Se4 | 0.506 | 0 | 4.125 | 0.079199 |
| **JVASP-60098** | YFeO3 | 0.805 | 0 | 24.747 | 0.037703 |
| **JVASP-8096** | Cr2FeSe4 | 0.613 | 0 | 5.997 | 0 |
| **JVASP-58040** | Ba2TlFe2O7 | 1.942 | 0 | 7.098 | 0 |
| **JVASP-54955** | KMnAg3C6N6 | 1.008 | 0 | 1 | 0.166123 |
| **JVASP-14753** | HfZn2 | 1.052 | 0 | 1.177 | 0.002744 |
| **JVASP-55644** | KFeBr3 | 0.339 | 0 | 16 | 0 |
| **JVASP-38339** | RbAuO3 | 1.701 | 0 | 1.329 | 0.414073 |
| **JVASP-58042** | Ba2TlCo2O7 | 0.317 | 0 | 5.352 | 0 |
| **JVASP-8618** | CrPbO3 | 1.021 | 0 | 2 | 0 |
| **JVASP-5314** | UCl5 | 1.152 | 0.41 | 2 | 0 |
| **JVASP-59564** | V2CdO4 | 1.025 | 0 | 8 | 0 |
| **JVASP-58337** | KY2Ti2S2O5 | 0.326 | 0 | 0.917 | 0 |
| **JVASP-59847** | YMn3Se2ClO8 | 0.419 | 0 | 18.612 | 0 |
| **JVASP-60099** | YCoO3 | 1.365 | 0 | 11.449 | 0.139314 |
| **JVASP-36101** | Er2Co3Ge5 | 1.248 | 0 | 0.916 | 0 |
| **JVASP-57577** | TiFeBi2O6 | 0.284 | 0 | 4.001 | 0 |
| **JVASP-56803** | Ba3Fe3Se7 | 0.484 | 0 | 19.803 | 0 |
| **JVASP-8619** | SrCrO3 | 0.639 | 0 | 1.973 | 0 |
| **JVASP-57326** | Ni2Ag3O4 | 1.039 | 0 | 2.738 | 0.02373 |
| **JVASP-56962** | CrGaCo2 | 1.008 | 0 | 3.035 | 0 |
| **JVASP-58227** | BaMgCo4O8 | 0.411 | 0 | 11.524 | 0 |
| **JVASP-7958** | MnAsRh | 0.32 | 0 | 8.557 | 0.087241 |
| **JVASP-59570** | Ba3Cr2O8 | 0.409 | 0 | 2 | 0 |
| **JVASP-8018** | MnIr | 2.046 | 0 | 2.495 | 0 |
| **JVASP-37845** | CuSeO4 | 0.455 | 0 | 1.95 | 6.24E-05 |
| **JVASP-58095** | YCoO3 | 1.009 | 0 | 3.761 | 0.163071 |
| **JVASP-7892** | BaY2NiO5 | 1.009 | 0 | 1.252 | 0 |
| **JVASP-57868** | CrSn2 | 0.29 | 0 | 1.647 | 0.181985 |
| **JVASP-60101** | YMoO3 | 0.837 | 0 | 10.18 | 0.053661 |
| **JVASP-12919** | CuPtF6 | 0.631 | 0 | 1.999 | 0 |
| **JVASP-59682** | RbHgN3O6 | 1.012 | 0 | 4 | 0.343516 |
| **JVASP-56072** | MnBi | 2.039 | 0 | 6.924 | 0.277582 |
| **JVASP-21191** | Ca2NiIrO6 | 1.31 | 0 | 1.549 | 0 |
| **JVASP-59496** | Cd2Re2O7 | 2.491 | 0 | 1.746 | 0 |
| **JVASP-42985** | Li5Nb2V5O2 | 1.047 | 0 | 3.417 | 0 |
| **JVASP-59573** | Sr3NiIrO6 | 2.902 | 0 | 4.471 | 0 |
| **JVASP-59637** | Mn4Ge6Ir7 | 1.728 | 0 | 16.86 | 0 |
| **JVASP-41091** | Ta2FeOs | 0.945 | 0 | 1.545 | 0 |
| **JVASP-55190** | Rb2FeI4 | 0.373 | 0 | 8 | 0.000214 |
| **JVASP-60102** | YVO3 | 1.014 | 0 | 10.626 | 0.087396 |
| **JVASP-8623** | FeBiO3 | 0.323 | 0 | 3.258 | 0.041491 |
| **JVASP-56304** | TlCoCl3 | 1.032 | 0 | 5.985 | 0 |
| **JVASP-21588** | Sr2ScIrO6 | 3.617 | 0 | 3.152 | 0 |
| **JVASP-58461** | BaFe4O8 | 0.266 | 0 | 10.253 | 0 |
| **JVASP-2817** | K4IrO4 | 0.632 | 0.37 | 2.88 | 0 |
| **JVASP-14308** | Ta2InCuTe4 | 0.508 | 0 | 2.969 | 0.336609 |
| **JVASP-13174** | Ta2CrO6 | 0.451 | 0 | 7.356 | 0 |
| **JVASP-56818** | MnSbPt | 0.854 | 0 | 3.913 | 0.372015 |
| **JVASP-59580** | BaVO3 | 0.653 | 0 | 1.154 | 0 |
| **JVASP-49659** | Nb2Co2O9 | 1.003 | 0 | 3.882 | 0.15346 |
| **JVASP-8538** | CeNiSb2 | 2.223 | 0 | 0.546 | 0 |
| **JVASP-58250** | YCo4B | 0.544 | 0 | 6.681 | 0 |
| **JVASP-59704** | Fe2CuGe2 | 0.393 | 0 | 0.508 | 0 |
| **JVASP-59773** | Mn3GeIr | 0.859 | 0 | 38.663 | 0.138852 |
| **JVASP-54797** | CrCoPt2 | 1.687 | 0 | 1.199 | 0 |
| **JVASP-58411** | Rb2CoSe2 | 1.034 | 0 | 5.996 | 0 |
| **JVASP-59709** | In2CoS4 | 0.547 | 0 | 5.991 | 0 |
| **JVASP-55805** | LiMnAsO4 | 8.433 | 0.09 | 20 | 0 |
| **JVASP-59509** | Ta3Co3C | 1.468 | 0 | 4.005 | 0 |
| **JVASP-59587** | YMnO3 | 1.006 | 0 | 7.998 | 0 |
| **JVASP-8240** | YFeO3 | 0.618 | 0 | 2.557 | 0.197367 |
| **JVASP-41131** | Hf2CoIr | 1.023 | 0 | 0.69 | 0 |
| **JVASP-59648** | Mn3SiIr | 2.239 | 0 | 25.834 | 0.052054 |
| **JVASP-21327** | Zn2BiWO6 | 1.488 | 0 | 1.15 | 0 |
| **JVASP-20932** | Cr2CuSe4 | 0.45 | 0 | 10.336 | 0 |
| **JVASP-58257** | FeRh2S4 | 1.214 | 0 | 6.105 | 0 |
| **JVASP-7856** | YCrO3 | 1.228 | 0 | 2.978 | 0.144098 |
| **JVASP-59882** | YCo2S4 | 3.012 | 0 | 1.785 | 0.142833 |
| **JVASP-21582** | MnNb3S6 | 1.016 | 0 | 8.337 | 0 |
| **JVASP-57296** | SrFe2Se4O2 | 1.933 | 0 | 5.999 | 0 |
| **JVASP-8323** | CoBi2O6 | 2.093 | 0 | 1.784 | 0.180329 |
| **JVASP-59593** | NiRh2O4 | 0.322 | 0 | 3.986 | 0 |
| **JVASP-20447** | FePt | 0.413 | 0 | 3.243 | 0 |
| **JVASP-38273** | KRhO3 | 0.349 | 0 | 1.784 | 0 |
| **JVASP-21693** | HfMn2 | 2.318 | 0 | 2.938 | 0 |
| **JVASP-45533** | YFeO3 | 1.32 | 0 | 11.999 | 0 |
| **JVASP-7858** | TlFeO2 | 1.084 | 0 | 5.29 | 0 |
| **JVASP-21125** | Sr3LiIrO6 | 4.647 | 0.2 | 3.869 | 0 |
| **JVASP-21071** | Cr2HgSe4 | 1.029 | 0 | 11.956 | 0 |
| **JVASP-20414** | Co3W | 0.255 | 0 | 0.833 | 0.096612 |
| **JVASP-26876** | CeH2 | 0.973 | 0 | 0.763 | 0 |
| **JVASP-38639** | Ni3Au | 1.081 | 0 | 1.538 | 0.088389 |
| **JVASP-21589** | Ba2ScIrO6 | 2.197 | 0 | 3.073 | 0 |
| **JVASP-59107** | Zr6Co23 | 0.415 | 0 | 25.261 | 0 |
| **JVASP-41448** | TmUTc2 | 2.133 | 0 | 0.877 | 0 |
| **JVASP-20397** | ZrMn2 | 0.264 | 0 | 2.878 | 0.000832 |
| **JVASP-8633** | VPbO3 | 0.328 | 0 | 1 | 0 |
| **JVASP-8737** | MnPd3 | 0.372 | 0 | 9.241 | 0 |
| **JVASP-26681** | Na3OsO5 | 1.26 | 0.18 | 2.702 | 0 |
| **JVASP-8433** | BaMnGe | 0.329 | 0 | 7.109 | 0 |
| **JVASP-12965** | Mn3As2 | 0.392 | 0 | 14.986 | 0.095392 |
| **JVASP-26231** | Co2B6Mo2 | 0.444 | 0 | 17.539 | 0 |
| **JVASP-20780** | Rb2O3 | 0.942 | 0 | 4 | 0 |
| **JVASP-8249** | YCrO3 | 1.23 | 0 | 2.978 | 0.143851 |
| **JVASP-39263** | FeAg3 | 0.397 | 0 | 2.97 | 0.356336 |
| **JVASP-24357** | Sr3CaIrO6 | 2.334 | 0 | 1.645 | 0 |
| **JVASP-41360** | HfScCo2 | 1.189 | 0 | 0.652 | 0 |
| **JVASP-22381** | Sr4IrO6 | 2.288 | 0 | 1.604 | 0 |
| **JVASP-22441** | Na3Cd2IrO6 | 1.503 | 0.23 | 1.831 | 0 |
| **JVASP-22442** | Ba3NaIrO6 | 1.853 | 0.16 | 3.946 | 0 |
| **JVASP-49567** | MgMo6O6 | 2.044 | 0 | 2.361 | 0 |
| **JVASP-39364** | Ni3Hg | 1.068 | 0 | 0.883 | 0.285456 |
| **JVASP-25704** | NbVF6 | 0.466 | 0 | 4 | 0 |
| **JVASP-15904** | MnSbIr | 1.029 | 0 | 3.103 | 0 |
| **JVASP-8204** | NiPt | 2.669 | 0 | 1.828 | 2.00E-05 |
| **JVASP-26047** | Co2Re2B6 | 0.768 | 0 | 17.729 | 0 |
| **JVASP-8335** | ZnCr2N2 | 0.288 | 0 | 5.855 | 0.382623 |
| **JVASP-39275** | FeAu3 | 0.315 | 0 | 3.117 | 0.168094 |
| **JVASP-8122** | NaRuO2 | 0.496 | 0.11 | 1 | 0.021784 |
| **JVASP-26796** | Li2MoF6 | 1.034 | 0 | 4 | 0 |
| **JVASP-7922** | BaMn2P2 | 0.342 | 0 | 2.791 | 0 |
| **JVASP-7868** | FeAgO2 | 0.766 | 0 | 2.083 | 0 |
| **JVASP-38211** | Rb3Pb | 1.125 | 0 | 0.912 | 0 |
| **JVASP-38310** | RbInO3 | 0.298 | 0 | 2 | 0.284021 |
| **JVASP-39279** | FePbO3 | 0.298 | 0 | 3.441 | 0 |
| **JVASP-7869** | Sr2MnO4 | 0.283 | 0 | 3 | 0 |
| **JVASP-26074** | Rb3Fe2Se4 | 1.094 | 0 | 30.015 | 0 |
| **JVASP-16563** | MnSnIr | 0.672 | 0 | 3.447 | 0.344516 |
| **JVASP-7702** | TiAu | 1.023 | 0 | 0.934 | 0.113969 |
| **JVASP-38264** | MnTl3 | 1.228 | 0 | 8.367 | 0.456432 |
| **JVASP-38317** | RbF3 | 0.411 | 0 | 2 | 0.161013 |
| **JVASP-20801** | Cr2Te3 | 1.454 | 0 | 24.218 | 0 |
| **JVASP-21389** | Sr3CuPtO6 | 2.174 | 0.21 | 1.919 | 0 |
| **JVASP-14273** | TiCdHg2 | 0.529 | 0 | 0.835 | 0.175958 |
| **JVASP-15949** | ZnFe3C | 0.295 | 0 | 4.078 | 0 |
| **JVASP-38319** | RbMnO3 | 0.387 | 0 | 1.908 | 0 |
| **JVASP-22401** | Sr3MnN3 | 1.01 | 0 | 2.486 | 0 |
| **JVASP-38492** | K3Rh | 1.073 | 0 | 0.501 | 0.477367 |
| **JVASP-19159** | YFe4Cu3O2 | 0.819 | 0 | 9.021 | 0 |
| **JVASP-15951** | ZnCo3C | 1.005 | 0 | 0.927 | 0 |
| **JVASP-8510** | Cr2As | 0.86 | 0 | 3.255 | 0.162059 |
| **JVASP-39287** | K3Sn | 0.424 | 0 | 0.912 | 0 |
| **JVASP-38226** | MnSn3 | 0.302 | 0 | 7.979 | 0.339145 |
| **JVASP-21417** | Sr3ZnIrO6 | 2.193 | 0.07 | 1.436 | 0 |
| **JVASP-38325** | RbNiO3 | 0.383 | 0 | 2.666 | 0 |
| **JVASP-26526** | Ba2Fe2S2OF2 | 0.314 | 0 | 7.827 | 0 |
| **JVASP-8301** | MgMoF6 | 1.013 | 0 | 2 | 0 |
| **JVASP-15918** | LiTiTe2 | 1.022 | 0 | 0.843 | 0 |
| **JVASP-26527** | SrCrF6 | 0.284 | 0 | 2 | 0 |
| **JVASP-21422** | Ba2CaOsO6 | 0.603 | 0 | 1.822 | 0 |
| **JVASP-38183** | Rb3Tl | 1.317 | 0 | 1.175 | 0.035357 |
| **JVASP-8567** | Li2FeBr4 | 0.852 | 0 | 4 | 0 |
| **JVASP-8609** | Sr2CoCl2O2 | 0.414 | 0 | 2.115 | 0.000768 |
| **JVASP-16654** | CoCu2SnSe4 | 0.616 | 0 | 2.458 | 0 |
| **JVASP-26528** | Rb2IrF6 | 1.404 | 0.35 | 0.996 | 0 |
| **JVASP-27213** | Ni2Hg2OF6 | 1.058 | 0 | 8.165 | 0 |
| **JVASP-37387** | Th3U | 2.582 | 0 | 2.336 | 0.250054 |
| **JVASP-15391** | TlCo2Se2 | 0.426 | 0 | 1.794 | 0 |
| **JVASP-16407** | LiRhF6 | 0.561 | 0 | 1.999 | 0 |
| **JVASP-21502** | Li2RhF6 | 2.022 | 0.06 | 1.993 | 0 |
| **JVASP-16367** | MnAlPt | 1.054 | 0 | 6.85 | 0 |
| **JVASP-27661** | Zn2Fe3O8 | 0.465 | 0 | 12 | 0 |
| **JVASP-38614** | MgPbO3 | 0.391 | 0 | 2 | 0.092507 |
| **JVASP-16368** | MnAlPt2 | 1.057 | 0 | 4.176 | 0 |
| **JVASP-38289** | RbSrO3 | 0.331 | 0 | 3 | 0 |
| **JVASP-16408** | LiIrF6 | 1.158 | 0.28 | 2 | 0 |
| **JVASP-26823** | Ba3NiIr2O9 | 3.343 | 0 | 9.097 | 0 |
| **JVASP-19598** | V3Te4 | 0.92 | 0 | 6.915 | 0.008259 |
| **JVASP-16409** | K2IrF6 | 1.396 | 0.3 | 0.995 | 0 |
| **JVASP-22415** | Sr3MgIrO6 | 1.99 | 0.12 | 1.588 | 0 |
| **JVASP-16823** | CrTe4Au | 0.542 | 0 | 2.872 | 0 |
| **JVASP-14403** | Mn2Sb | 0.5 | 0 | 11.488 | 0.333028 |
| **JVASP-38341** | RbAgO3 | 1.28 | 0 | 1.663 | 0.39356 |
| **JVASP-8383** | YWF5 | 1.5 | 0 | 3.498 | 0 |
| **JVASP-17460** | Ba2Mn3As2O2 | 3.292 | 0 | 3.371 | 0 |
| **JVASP-858** | Co | 0.531 | 0 | 3.153 | 5.10E-06 |
| **JVASP-55303** | Mn3TeO6 | 0.305 | 0 | 29.848 | 0.054318 |
| **JVASP-17641** | CaMnGe | 0.313 | 0 | 4.563 | 0 |
| **JVASP-55470** | Co2Mo3O8 | 7.451 | 0 | 0.623 | 0 |
| **JVASP-19739** | TiHg | 1.407 | 0 | 1.372 | 0.148755 |
| **JVASP-17315** | CrAsRh | 0.259 | 0 | 10.625 | 0.000782 |
| **JVASP-8384** | YNiF5 | 0.372 | 0.01 | 2 | 0 |
| **JVASP-38244** | Rb3In | 0.418 | 0 | 1.284 | 0.094258 |
| **JVASP-39452** | RuAu3 | 1.302 | 0 | 1.719 | 0.368509 |
| **JVASP-17265** | BaIrF6 | 1.356 | 0.39 | 0.997 | 0 |
| **JVASP-17509** | RhO2F6 | 3.993 | 0 | 4.048 | 4.00E-06 |
| **JVASP-8349** | ZnCoF6 | 0.321 | 0 | 1 | 0 |
| **JVASP-19704** | NbF4 | 0.254 | 0 | 0.682 | 0 |
| **JVASP-17643** | CoNiSn | 1.014 | 0 | 1.788 | 0.007715 |
| **JVASP-39453** | RuAu3 | 1.487 | 0 | 3.489 | 0.397432 |
| **JVASP-38515** | KAgO3 | 0.91 | 0 | 1.549 | 0.394317 |
| **JVASP-37981** | CoTeO3 | 0.544 | 0 | 0.729 | 0.307307 |
| **JVASP-16338** | FeTe | 0.654 | 0 | 3.87 | 0.207063 |
| **JVASP-37453** | TaTiFe2 | 1.042 | 0 | 0.83 | 0 |
| **JVASP-8385** | YCoF5 | 1.175 | 0.2 | 3 | 0 |
| **JVASP-12407** | TiFe6Ge6 | 1.016 | 0 | 8.704 | 0 |
| **JVASP-38248** | Rb3Ga | 0.469 | 0 | 1.472 | 0.112694 |
| **JVASP-17730** | FeSn | 1.009 | 0 | 3.584 | 0.176643 |
| **JVASP-16719** | YFe2B2 | 0.429 | 0 | 1.027 | 0 |
| **JVASP-47356** | Li2Si2WO7 | 0.887 | 0 | 4 | 0 |
| **JVASP-12287** | ZrMnGe | 0.267 | 0 | 8.375 | 0 |
| **JVASP-12608** | Sr3Co2Cl2O5 | 0.693 | 0 | 3.49 | 0 |
| **JVASP-17854** | Mn3SnC | 0.392 | 0 | 3.33 | 0.009175 |
| **JVASP-17646** | RbMnAs | 0.255 | 0 | 3.881 | 0 |
| **JVASP-37314** | SrAlO3 | 2.995 | 0 | 0.526 | 0 |
| **JVASP-36857** | MnAuO2 | 0.374 | 0 | 3.984 | 0 |
| **JVASP-46728** | Li4Mn3SbP4O6 | 3.224 | 0.09 | 15.761 | 0 |
| **JVASP-21188** | Ca2FeIrO6 | 0.982 | 0 | 5.586 | 0 |
| **JVASP-17462** | Sr2Mn3As2O2 | 3.371 | 0 | 7.453 | 0 |
| **JVASP-18123** | FePd3 | 0.402 | 0 | 4.169 | 0 |
| **JVASP-19789** | CrSb | 0.378 | 0 | 5.604 | 0.153325 |
| **JVASP-17168** | MoPt3 | 1.132 | 0 | 1.472 | 0.15261 |
| **JVASP-37028** | TiInFe2 | 2.041 | 0 | 0.907 | 0 |
| **JVASP-16723** | CrIr3 | 1.093 | 0 | 1.237 | 0 |
| **JVASP-10854** | YFe2O4 | 0.834 | 0 | 14 | 0.213896 |
| **JVASP-37407** | TePdO3 | 0.99 | 0 | 1.019 | 0 |
| **JVASP-37144** | NbFe3 | 1.008 | 0 | 5.132 | 0 |
| **JVASP-49613** | Y2Co2O7 | 0.266 | 0 | 1.903 | 0 |
| **JVASP-44414** | Li2Mn3WO8 | 1.054 | 0 | 16.292 | 0 |
| **JVASP-46240** | CoBi2O6 | 4.185 | 0 | 0.549 | 0 |
| **JVASP-12612** | Ba2UCoO6 | 1.058 | 0 | 3.011 | 0 |
| **JVASP-18049** | CeB6 | 0.88 | 0 | 0.721 | 0 |
| **JVASP-15970** | Cr3SnN | 1.006 | 0 | 1.687 | 0 |
| **JVASP-16474** | TiCdHg2 | 0.529 | 0 | 0.835 | 0.175958 |
| **JVASP-44705** | MnSb4O2 | 1.432 | 0.02 | 3 | 0 |
| **JVASP-10855** | ZnFe2O4 | 3.993 | 0 | 4.001 | 5.50E-05 |
| **JVASP-37189** | Mn3Ga | 0.754 | 0 | 3.11 | 0.095038 |
| **JVASP-11508** | TiBi2O6 | 1.145 | 0 | 3.857 | 0.332215 |
| **JVASP-11584** | Sr4MgFe2S2O6 | 0.343 | 0 | 8.37 | 0 |
| **JVASP-37701** | Y3Sn | 0.286 | 0 | 1.596 | 0.106332 |
| **JVASP-8357** | AlWF5 | 1.143 | 0 | 3.97 | 0 |
| **JVASP-8414** | BaYCoCuO5 | 0.557 | 0 | 2.252 | 0 |
| **JVASP-49615** | YMoO3 | 0.705 | 0 | 3.257 | 0.001451 |
| **JVASP-44418** | Na3CrBAsO7 | 6.077 | 1.09 | 6 | 0 |
| **JVASP-17654** | MnGaPt | 0.463 | 0 | 3.336 | 0.246664 |
| **JVASP-37190** | MnGaFeCo | 1.018 | 0 | 3.068 | 0 |
| **JVASP-17478** | Sc3In | 1.013 | 0 | 2.465 | 0 |
| **JVASP-45802** | Li8Cr3TeO2 | 3.222 | 0 | 5.932 | 0 |
| **JVASP-27678** | Mn4ZnCu3O2 | 1.061 | 0 | 9.079 | 0 |
| **JVASP-18328** | NaMnBi | 0.348 | 0 | 8.108 | 0.057047 |
| **JVASP-11811** | Ba2Mn2Sb2O | 0.444 | 0 | 19.229 | 0 |
| **JVASP-18131** | FeCu2Sn | 0.881 | 0 | 2.688 | 0.263135 |
| **JVASP-12355** | Zr2Fe3Ge | 0.336 | 0 | 7.794 | 0 |
| **JVASP-38645** | Ni3Au | 1.248 | 0 | 3.103 | 0.097749 |
| **JVASP-11692** | RbFeMo2O8 | 0.7 | 0 | 1 | 0 |
| **JVASP-37208** | SiPdO3 | 1.016 | 0 | 1.954 | 0 |
| **JVASP-44507** | Li2MnBAsO7 | 2.041 | 1.17 | 6 | 0 |
| **JVASP-10857** | ZnCr2Se4 | 1.012 | 0 | 11.98 | 0 |
| **JVASP-17618** | Mn2GaCo | 2.034 | 0 | 2.019 | 0 |
| **JVASP-37946** | Co3Bi | 1.067 | 0 | 2.899 | 0.443311 |
| **JVASP-18209** | Mn3Sn | 0.789 | 0 | 10.047 | 0.202535 |
| **JVASP-18104** | VCo2Sn | 0.411 | 0 | 2.813 | 0 |
| **JVASP-17328** | Mn3ZnN | 1.005 | 0 | 4.201 | 0 |
| **JVASP-38764** | ZnFeRh2 | 0.52 | 0 | 4.221 | 0 |
| **JVASP-52119** | MnReO4 | 2.649 | 0 | 7.645 | 1.75E-06 |
| **JVASP-17828** | ZnFeSb | 0.582 | 0 | 2.538 | 0.474698 |
| **JVASP-18136** | NdCoSi | 0.936 | 0 | 0.511 | 0 |
| **JVASP-37600** | Sr3Cr | 1.006 | 0 | 4.864 | 0.495398 |
| **JVASP-8416** | BaYVCuO5 | 3.021 | 0 | 1.571 | 0 |
| **JVASP-44720** | P2WO7 | 2.354 | 0 | 4 | 0.008388 |
| **JVASP-37040** | TiAu | 1.722 | 0 | 1.708 | 0.04604 |
| **JVASP-18366** | K2RuCl6 | 1.011 | 0 | 2 | 0 |
| **JVASP-42952** | Li4Co3TeO8 | 0.945 | 0 | 9.012 | 0 |
| **JVASP-19792** | Fe3Pt | 1.068 | 0 | 8.229 | 0.057277 |
| **JVASP-46260** | Li2Ni3BiO8 | 2.085 | 0 | 2.988 | 0 |
| **JVASP-37424** | Ta2Be2O5 | 1.309 | 0 | 0.702 | 0 |
| **JVASP-10656** | ZnMo2O4 | 0.664 | 0 | 3.853 | 0.156794 |
| **JVASP-36884** | TlFeF3 | 0.305 | 0 | 4 | 0 |
| **JVASP-16803** | MnAu | 0.38 | 0 | 4.067 | 0.121371 |
| **JVASP-44512** | Li4Mn5NbO2 | 0.993 | 0 | 2.81 | 0 |
| **JVASP-44654** | LiZnFe0O6 | 8.443 | 0 | 15.74 | 0 |
| **JVASP-16853** | Ni3Pt | 1.307 | 0 | 2.199 | 0 |
| **JVASP-16804** | MnAu | 1.04 | 0 | 4.069 | 0.121112 |
| **JVASP-36885** | TlCoF3 | 0.297 | 0 | 3 | 0 |
| **JVASP-17300** | BaMn2As2 | 0.35 | 0 | 3.707 | 0 |
| **JVASP-17454** | Sr2CoO4 | 0.271 | 0 | 1.953 | 0 |
| **JVASP-18213** | MnCu2SnSe4 | 0.296 | 0 | 4.636 | 0 |
| **JVASP-16854** | CoCu2Sn | 0.4 | 0 | 0.955 | 0.222465 |
| **JVASP-17624** | PuFe2Si2 | 2.996 | 0 | 4.569 | 0 |
| **JVASP-18302** | VGaCo2 | 0.415 | 0 | 1.943 | 5.55E-05 |
| **JVASP-16806** | NaMnTe2 | 1.044 | 0 | 4.052 | 0.093483 |
| **JVASP-18368** | K2OsCl6 | 1.079 | 0.02 | 1.999 | 0 |
| **JVASP-11590** | Sr4CaFe2S2O6 | 0.482 | 0 | 8.477 | 0 |
| **JVASP-37044** | TiFe2As | 0.274 | 0 | 1.008 | 0 |
| **JVASP-45852** | Li2Nb2Fe3O0 | 1.282 | 0.08 | 8.067 | 0 |
| **JVASP-17625** | Mn3GeC | 0.333 | 0 | 2.997 | 0 |
| **JVASP-44518** | Li5Nb2Fe5O2 | 5.969 | 0 | 15 | 0.043384 |
| **JVASP-18303** | MnTePd | 0.49 | 0 | 4.838 | 0 |
| **JVASP-20640** | FePt | 0.413 | 0 | 3.243 | 0 |
| **JVASP-17790** | Mn2Sb | 1.03 | 0 | 2.746 | 0.124949 |
| **JVASP-17237** | FeNiPt2 | 0.661 | 0 | 4.629 | 0 |
| **JVASP-18214** | BaMn2Ge2 | 1.008 | 0 | 4.419 | 0 |
| **JVASP-34926** | Sr5Bi3 | 1.492 | 0 | 1.639 | 0 |
| **JVASP-17471** | Sr2Mn3Sb2O2 | 3.303 | 0 | 8.403 | 0 |
| **JVASP-36950** | CaTcO3 | 0.45 | 0 | 1.819 | 0 |
| **JVASP-9317** | YFeW2O8 | 4.021 | 0 | 3.593 | 0 |
| **JVASP-18255** | InFe2CuSe4 | 1.021 | 0 | 7.611 | 0 |
| **JVASP-46173** | HfFeO3 | 1.518 | 0 | 16 | 0 |
| **JVASP-18349** | TiI3 | 0.969 | 0 | 0.704 | 0.003149 |
| **JVASP-45854** | Li3Nb2Fe3O0 | 2.947 | 0 | 5.002 | 0.026501 |
| **JVASP-17457** | Ba2Mn3Sb2O2 | 1.028 | 0 | 9.098 | 0 |
| **JVASP-11697** | LiFeAs2O7 | 0.353 | 0 | 4.997 | 0 |
| **JVASP-17394** | MnCu2Sb | 0.752 | 0 | 3.86 | 0.300664 |
| **JVASP-18174** | Co2As | 1.064 | 0 | 4.734 | 0.27925 |
| **JVASP-44438** | Li2Co3SnO8 | 0.326 | 0 | 1.631 | 0 |
| **JVASP-17718** | Mn2Ge | 0.344 | 0 | 8.937 | 0.088172 |
| **JVASP-19978** | NbF3 | 0.308 | 0 | 2 | 0.08838 |
| **JVASP-37051** | Ti2GaFe | 0.856 | 0 | 0.982 | 0 |
| **JVASP-44593** | Li4Mn3Sn5O6 | 3.986 | 0 | 12.94 | 0 |
| **JVASP-19388** | Ca2FeSbO6 | 0.337 | 0 | 8.905 | 0 |
| **JVASP-44528** | Li2Co3TeO8 | 1.45 | 0 | 0.962 | 0 |
| **JVASP-11093** | ZnFe4S8 | 0.621 | 0 | 6.639 | 0.033962 |
| **JVASP-34927** | Ba5Bi3 | 2.017 | 0 | 1.848 | 0 |
| **JVASP-11592** | Sr4MgCo2S2O6 | 1.067 | 0 | 5.16 | 0 |
| **JVASP-17303** | CdRhF6 | 0.767 | 0.09 | 0.992 | 0 |
| **JVASP-17637** | MnGaNi2 | 0.298 | 0 | 4.013 | 0.00848 |
| **JVASP-45210** | NbV3O8 | 4.062 | 0 | 7.962 | 0 |
| **JVASP-18176** | Rb2RhF6 | 1.013 | 0.19 | 0.996 | 0 |
| **JVASP-46361** | TaFeO4 | 2.096 | 0 | 4.012 | 0 |
| **JVASP-34315** | CuSeO4 | 0.457 | 0 | 1.95 | 0 |
| **JVASP-20585** | ZrMn2 | 0.264 | 0 | 2.878 | 0.000832 |
| **JVASP-17721** | MnSnPd2 | 0.25 | 0 | 4.136 | 0 |
| **JVASP-9209** | Ba2YCo3O7 | 1.016 | 0 | 6.49 | 0 |
| **JVASP-18028** | FeTe | 0.283 | 0 | 4.155 | 0.104371 |
| **JVASP-36838** | NiAuO2 | 0.432 | 0 | 1.669 | 0 |
| **JVASP-18220** | Mn2CoSn | 0.357 | 0 | 1.885 | 0.150639 |
| **JVASP-8486** | PuNi5 | 3.485 | 0 | 3.303 | 0 |
| **JVASP-12626** | InFeO3 | 0.624 | 0 | 5.654 | 0 |
| **JVASP-20602** | Co3W | 0.255 | 0 | 0.833 | 0.096613 |
| **JVASP-17304** | HgRhF6 | 0.701 | 0 | 0.981 | 0 |
| **JVASP-44617** | Li2Fe3SnO8 | 0.503 | 0 | 4.156 | 0.085692 |
| **JVASP-44742** | Li5Ni5Sn2O2 | 5.359 | 0.02 | 8.956 | 0 |
| **JVASP-46857** | Li2Co3SnO8 | 0.27 | 0 | 0.859 | 0.020608 |
| **JVASP-44461** | VBiO3 | 0.756 | 0 | 8 | 0.014917 |
| **JVASP-19396** | Ca2SbMoO6 | 0.43 | 0 | 2.121 | 0 |
| **JVASP-12627** | Sr3Fe2Cl2O5 | 0.288 | 0 | 7.522 | 0 |
| **JVASP-18082** | MnSnPt | 0.399 | 0 | 3.63 | 0 |
| **JVASP-17803** | FeSb2 | 0.727 | 0 | 1.72 | 0 |
| **JVASP-46861** | FeSbO4 | 0.268 | 0 | 5.77 | 0 |
| **JVASP-42911** | LiMnSbO4 | 4.231 | 0 | 9.477 | 0 |
| **JVASP-10907** | AlBi3O9 | 0.588 | 0 | 5.951 | 0.027749 |
| **JVASP-18193** | FeCo2Ge | 0.291 | 0 | 5.279 | 0 |
| **JVASP-43066** | CoSbO4 | 0.357 | 0 | 0.776 | 0 |
| **JVASP-44991** | FeSb4O2 | 2.071 | 0.04 | 2 | 0 |
| **JVASP-44750** | Li2Ni3WO8 | 1.136 | 0 | 3.997 | 0 |
| **JVASP-34319** | TlV3Cr2S8 | 2.865 | 0 | 6.571 | 0 |
| **JVASP-45016** | LiFeSnO4 | 0.456 | 0 | 2 | 0 |
| **JVASP-12631** | CdFe2O4 | 1.621 | 0 | 4.006 | 0 |
| **JVASP-10484** | Ba2SrIrO6 | 0.824 | 0.13 | 2.617 | 0 |
| **JVASP-9343** | VW2O8 | 0.773 | 0 | 0.894 | 0 |
| **JVASP-9897** | Mg2CrWO6 | 1.216 | 0 | 3.683 | 0 |
| **JVASP-12217** | ZnFe2O4 | 6.588 | 0 | 4.001 | 4.93E-06 |
| **JVASP-9466** | Ba2TlNi2O7 | 0.465 | 0 | 3.151 | 0 |
| **JVASP-34389** | FeSnF6 | 0.503 | 0 | 4.039 | 0 |
| **JVASP-9216** | Ba2YNi3O7 | 0.288 | 0 | 0.603 | 0 |
| **JVASP-12231** | KFeMo2O8 | 0.529 | 0 | 1 | 0 |
| **JVASP-12637** | FeMoClO4 | 2.006 | 0 | 4.497 | 0 |
| **JVASP-11536** | YV2O4 | 0.506 | 0 | 9.995 | 0.094075 |
| **JVASP-10731** | ZnCo4O8 | 0.253 | 0 | 1.777 | 0.01025 |
| **JVASP-9469** | Ba2YTlV2O7 | 2.984 | 0 | 3.127 | 0 |
| **JVASP-9640** | YCu2O4 | 1.012 | 0 | 5.939 | 0.335261 |
| **JVASP-9527** | VZnSF5 | 1.25 | 0 | 3.998 | 0 |
| **JVASP-9203** | Ba2AlNi3O8 | 1.083 | 0 | 1.341 | 0 |
| **JVASP-9676** | ZnCr2Se4 | 1.012 | 0 | 11.98 | 1.06E-05 |
| **JVASP-9196** | Ba2AlCr3O8 | 3.056 | 0 | 9 | 0 |
| **JVASP-34404** | NbCrF6 | 3.008 | 0 | 2.115 | 0 |
| **JVASP-52140** | Ba3Ti3O8 | 0.38 | 0 | 1.571 | 0.016329 |
| **JVASP-34347** | BaFeF4 | 0.924 | 0 | 8 | 0 |
| **JVASP-9363** | BaZnFe4O8 | 1.021 | 0 | 18.898 | 0 |
| **JVASP-11670** | Sr2CoMoO6 | 0.312 | 0 | 2.993 | 0 |
| **JVASP-9258** | Sr2AlTlV2O7 | 3.397 | 0 | 4 | 0 |
| **JVASP-11731** | MnPt3O6 | 0.276 | 0 | 4.87 | 0 |
| **JVASP-9962** | V2ZnO4 | 0.342 | 0 | 6.627 | 0.117976 |
| **JVASP-9185** | Ba2YV3O8 | 1.012 | 0 | 2.24 | 0 |
| **JVASP-12643** | Ba2UMnO6 | 1.931 | 0 | 5 | 0 |
| **JVASP-34878** | K2Zr7Cl8 | 0.894 | 0 | 1.885 | 0 |
| **JVASP-34479** | Ba4Fe2S4I5 | 0.512 | 0 | 0.957 | 0 |
| **JVASP-10839** | YMnO3 | 1.006 | 0 | 7.998 | 1.11E-05 |
| **JVASP-9366** | BaCaCo4O8 | 0.802 | 0 | 7.42 | 0 |
| **JVASP-9966** | ZnFe2O4 | 0.529 | 0 | 12.329 | 0.144093 |
| **JVASP-34752** | KCdN3O6 | 0.544 | 0 | 4.001 | 0.377445 |
| **JVASP-12648** | MnSnB2O6 | 0.926 | 0.27 | 5 | 0 |
| **JVASP-34485** | ZrCoF6 | 0.875 | 0.18 | 3 | 0 |
| **JVASP-9267** | Sr2YTlV2O7 | 1.041 | 0 | 2.878 | 0 |
| **JVASP-24743** | Fe3W3N | 1.934 | 0 | 4.38 | 0 |
| **JVASP-34425** | Ba6Ru2PtCl2O2 | 4.102 | 0 | 5.99 | 0 |
| **JVASP-9187** | Ba2YCo3O8 | 1.025 | 0 | 2.726 | 0 |
| **JVASP-9269** | Ba2TlBi2O7 | 1.109 | 0 | 0.555 | 0 |
| **JVASP-78840** | Mn3Ge | 3.009 | 0 | 1.039 | 0.000129 |
| **JVASP-9491** | Sr2FeCuSO3 | 0.348 | 0 | 8.436 | 0 |
| **JVASP-9541** | YFeO3 | 1.005 | 0 | 8.072 | 0.059007 |
| **JVASP-34429** | ZrFeF6 | 1.004 | 0 | 4 | 0.013753 |
| **JVASP-10342** | Ge2MoO6 | 0.277 | 0 | 3.999 | 0 |
| **JVASP-52113** | Sr2CoReO6 | 0.407 | 0 | 2.047 | 0 |
| **JVASP-24596** | RbCoCl3 | 0.309 | 0 | 5.992 | 0 |
| **JVASP-9494** | ZnCoPO5 | 1.109 | 0 | 2.002 | 0 |
| **JVASP-79568** | MnGaFe2 | 1.942 | 0 | 2.075 | 0 |
| **JVASP-78684** | MnSnPt | 0.274 | 0 | 3.406 | 0.210315 |
| **JVASP-79239** | Mn3Pt | 0.791 | 0 | 1.908 | 0.093006 |
| **JVASP-9495** | Sr2CoSO3 | 1.033 | 0 | 4.155 | 0 |
| **JVASP-79241** | MnGePd2 | 0.428 | 0 | 4.112 | 0 |
| **JVASP-79574** | NbFe3 | 0.886 | 0 | 3.23 | 0.06218 |
| **JVASP-9931** | ZnFe2O4 | 5.142 | 0 | 4.001 | 0 |
| **JVASP-9496** | Sr2MnSO3 | 0.286 | 0 | 6.149 | 0 |
| **JVASP-79576** | MnGaFe2 | 0.319 | 0 | 6.399 | 0.037086 |
| **JVASP-78859** | NiBrO | 1.006 | 0 | 1.299 | 0 |
| **JVASP-79583** | Mn2CuGe | 0.438 | 0 | 0.669 | 0.107469 |
| **JVASP-79200** | VGaCo2 | 0.4 | 0 | 1.943 | 0 |
| **JVASP-24841** | Y6OsI0 | 1.144 | 0.04 | 1.177 | 0 |
| **JVASP-80740** | Ti2GaFe | 0.849 | 0 | 0.982 | 8.37E-05 |
| **JVASP-79431** | MgMnPt2 | 1.202 | 0 | 4.386 | 0 |
| **JVASP-79586** | VFeCoAs | 1.004 | 0 | 2.944 | 0 |
| **JVASP-16934** | YCoO3 | 1.005 | 0 | 1.171 | 0.25161 |
| **JVASP-78508** | CoSI | 0.685 | 0 | 1.969 | 0.44561 |
| **JVASP-78280** | Mn2Sb | 1.068 | 0 | 1 | 0.143192 |
| **JVASP-79206** | FeSe | 0.421 | 0 | 4.509 | 0.198133 |
| **JVASP-79916** | NbZnCo2 | 0.537 | 0 | 0.74 | 0 |
| **JVASP-78380** | BaN | 0.442 | 0 | 1 | 0.268514 |
| **JVASP-78429** | NaSe | 0.252 | 0 | 0.915 | 0.374696 |
| **JVASP-79435** | Mn2CoSn | 0.729 | 0 | 1.883 | 0.150843 |
| **JVASP-78470** | RbSe | 0.288 | 0 | 0.998 | 0.455465 |
| **JVASP-78658** | NiPt | 2.665 | 0 | 1.828 | 0 |
| **JVASP-79097** | Mg3Re | 1.082 | 0 | 0.51 | 0.361552 |
| **JVASP-79593** | Fe3Pt | 1.068 | 0 | 8.23 | 0.057144 |
| **JVASP-80098** | MnTe | 1.038 | 0 | 8.073 | 0.012304 |
| **JVASP-78434** | KSe | 0.252 | 0 | 0.996 | 0.421998 |
| **JVASP-79500** | VFeCoGe | 0.629 | 0 | 1.954 | 0 |
| **JVASP-82132** | MnBi2Te4 | 1.368 | 0.03 | 4.998 | 0 |
| **JVASP-80251** | CrSnRh2 | 0.297 | 0 | 2.89 | 0.238679 |
| **JVASP-78832** | MnBi | 2.039 | 0 | 6.925 | 0.277638 |
| **JVASP-79457** | CrInNi2 | 0.459 | 0 | 3.557 | 0.082176 |
| **JVASP-79604** | VGaFeCo | 0.413 | 0 | 0.952 | 0 |
| **JVASP-79562** | GaFeNi2 | 0.339 | 0 | 3.035 | 0.004254 |